\documentclass[11pt]{article}

\usepackage{multicol} 
\usepackage{fullpage}
\usepackage{amsmath}  
\usepackage{graphics}
\usepackage{url}
\usepackage[nottoc]{tocbibind} 

\newcommand{\surround}[1]{ \left(  #1 \right)  }
\newcommand{\surbrace}[1]{ \left\{ #1 \right\} }
\newcommand{\surbrack}[1]{ \left[  #1 \right]  }

\newcommand\Ateam{A^{\text{team}}}

\newcommand\Dteam{D^{\text{team}}}
\newcommand\Rteam{R^{\text{team}}}

\newcommand\Extract[1]{\surbrace{\bigstrut #1}}
\newcommand{\Eval}[1]{  \left.  #1 \right|  }
\newcommand{\Prob}{\text{Prob}}
\newcommand{\ds}{\displaystyle}

\newcommand\temp{}
\newcommand\ttemp{}

\newcommand\ZboxStrut[2]{
  \parbox{#1}{\raggedright \bigstrut #2 \bigstrut} 
}

\newcommand\Zbox[2]{
  \parbox{#1}{\raggedright #2 } 
}

\newcommand{\pdiff}[2]{     \frac{\partial     {#1}}{\partial {#2}}}

\newcommand\NotationDefinition[1]{ 
  \medskip
  \begin{center}
  \framebox{ \parbox{6.1in}{\raggedright #1} }
  \end{center}
  \medskip
}

\newcommand\kfrac[2]{\leavevmode\kern.1em
             \raise.50ex\hbox{\the\scriptfont0 #1}\kern-.1em
 /\kern-.15em\lower.25ex\hbox{\the\scriptfont0 #2}}

\newbox\bigstrutbox
\setbox\bigstrutbox=\hbox{\vrule height13pt depth 7pt width0pt}
\newcommand{\bigstrut}{\relax\ifmmode\copy\bigstrutbox%
                    \else\unhcopy\bigstrutbox\fi}

\newcommand{\smallspacing}{%
\setlength{\itemsep}{1pt}
\setlength{\listparindent}{0pt}
\setlength{\parsep}{0pt} 
\setlength{\parskip}{0pt}
\setlength{\partopsep}{0pt}
\setlength{\topsep}{0pt}
}

\usepackage{lastpage}                          
\usepackage{fancyhdr}                          
\begin{document}
%------------------------------------------------------
\title{Voting Power of Teams Working Together}
\author{Daniel Zwillinger \\\texttt{zwilling@az-tec.com}}
\maketitle

\begin{abstract}
\noindent
Voting power determines the ``power'' of individuals who cast votes;
their power is based on their ability to influence the winning-ness of
a coalition.
Usually each individual acts alone, casting either all or none of
their votes and is equally likely to do either.
This paper extends this standard ``random voting'' model to allow
probabilistic voting, partial voting, and correlated team voting.
We extend the standard Banzhaf metric to account for these cases; our
generalization reduces to the standard metric under ``random voting'',
This new paradigm allows us to answer questions such as ``In the 2013
US Senate, how much more unified would the Republicans have to be in
order to have the same power as the Democrats in attaining cloture?''

\end{abstract}


\bigskip
\bigskip
\tableofcontents

\newpage
%------------------------------------------------------
\section{Introduction}
%------------------------------------------------------
%
In a weighted voting game there are players who cast votes. 
Analysis of the players' voting power has a long history with
many models in use \cite{Felsenthal,LaruelleValenciano}.
The simplest voting model, ``random voting'', is when each voter is
equally likely to support or oppose a motion.  
Even in this simple case there are multiple ways to define the
players' power, the most common are the Shapley--Shubik power
\cite{ShapleyShubik} and the Banzhaf power
\cite{ComputationofPowerIndices,WikiBanzhafPowerIndex}.

Forming coalitions is a way for voters to influence their voting power
\cite{Gelman03formingvoting,GelmanKatzTuerlinckx}.  
Interestingly, Gelman \cite{Gelman03formingvoting} proves ``under the random voting
model, this average voting power is maximized under simple popular
vote (majority rule) and is lower under any coalition system'' and
makes the observation ``Joining a coalition is generally beneficial to
those inside the coalition but hurts those outside.''.
Not only can voters form coalitions, they can also vote
probabilistically.  
Some papers \cite{GelmanKatzTuerlinckx} create stochastic models for
coalitions of voters.

While using generating functions to compute the Shapley--Shubik power or
Banzhaf power  is well known, it has usually been in the
context of counting combinatorial possibilities for the ``random
voting'' model.
We have generalized the standard generating function approach to allow
more sophisticated models of voting to be analyzed.

The main results presented in this paper are the following:
\begin{enumerate} % \smallspacing

\item We review how Banzhaf power is defined and then illustrate the
  well-known process of determining Banzhaf power using generating
  functions for the ``random voting'' model.
  By generalizing to (simple) \textit{weighted} generating functions
  we show how to directly compute the Banzhaf power; we do not
  determine it indirectly via the usually obtained combinatorial
  counting.

\item Still using the ``random voting'' model we introduce
  \textit{Influence Polynomial}s; these are a proxy for a player's
  weighted generating function when used to compute the Banzhaf power.

\item We introduce a model of voting in which players have
  probabilities corresponding to the number of votes they cast.  
  This is represented by a (general) weighted generating function,
  which we call a \textit{voting structure}.
  We show how to determine players' Influence Polynomials from their
  voting structures.
  These Influence Polynomials allow a generalized Banzhaf power to be
  determined; this reduces to the usual Banzhaf power when the voting
  structure represents the ``random voting'' model.

\item We create and analyze voting structures for a coalition
  represented by a leader.
  In these coalitions each member follows the guidance of the leader
  probabilistically; not with certainty.

\end{enumerate}
Several examples are given, including an example related to the US
Senate.

%------------------------------------------------------
\section{Banzhaf power}
%------------------------------------------------------
In the usual way a \textit{weighted voting game} is represented by the
vector $[q; w_1 , w_2 , . . . , w_n]$ where:
\begin{enumerate} \smallspacing
\item There are $n$ players.
\item Player $i$ has $w_i$ votes (with $w_i > 0$).
\item A \textit{coalition} is a subset of players.
\item A coalition $S$ is \textit{winning} if $\sum_{i\in S} w_i \ge q$, where $q$ is the \textit{quota}.
\item A game is proper if $\frac{1}{2}\sum w_i < q$.
\end{enumerate}
To define the Banzhaf power consider all $2^n$ possible coalitions of
players.
For each coalition, if player~$i$ can change the winning-ness of the
coalition, by either entering or leaving the coalition, then player
$i$ is \textit{marginal}.  
The Banzhaf power index ($\beta$) of a player is proportional to the
number of times that a player is marginal; hence the total power of
all players is~1.

As a continuing example consider the $[6; 4, 3, 2, 1]$ weighted voting
game where the the players are named $\{A, B, C, D\}$ and ``random
voting'' is used.
There are $2^4=16$ subsets (or coalitions) of four players; the
following enumeration shows all coalitions (left) and the marginal
players for each (right):

\begin{multicols}{3}
\begin{enumerate} \smallspacing
\item $\{\,\} \to \{\,\} $
\item $\{A\} \to  \{B, C\}$
\item $\{B\} \to  \{A\}$
\item $\{C\} \to  \{A\}$
\item $\{D\} \to  \{\,\} $
\item $\{A, B\} \to  \{A, B\}$
\item $\{A, C\} \to  \{A, C\}$
\item $\{A, D\} \to  \{B, C\}$
\item $\{B, C\} \to  \{A, D\}$
\item $\{B, D\} \to  \{A, C\}$
\item $\{C, D\} \to  \{A, B\}$
\item $\{A, B, C\} \to  \{A\}$
\item $\{A, B, D\} \to  \{A, B\}$
\item $\{A, C, D\} \to  \{A, C\}$
\item $\{B, C, D\} \to  \{B, C, D\}$
\item $\{A, B, C, D\} \to  \{\,\} $
\end{enumerate}
\end{multicols}

Player $A$ is marginal 10 times, players $B$, $C$ are each marginal 6
times, and player $D$ is marginal 2 times. 
The total number of times that players are marginal is 24=10 + 6 + 6 + 2.
Hence player $A$ has Banzhaf power $\beta(A)=\frac{10}{24}=\frac{5}{12}$.
The other players have the powers:
$\beta(B)=\beta(C)=\frac{6}{24}=\frac{1}{4}$ and
$\beta(D)=\frac{2}{24}=\frac{1}{12}$. 
These powers can be determined by hand as shown above or using an
online tool such as \cite{OnlineComputation}.

% Note that the number of times each player is marginal is even. 
% %
% This is because coalitions in which a specific player is marginal are
% paired; one coalition has the player and one does not.

%------------------------------------------------------
\subsection{Banzhaf Power via Generating Functions}
%------------------------------------------------------
%
Imagine that each player in the $[6; 4, 3, 2, 1]$ game can choose,
with equal likelihood, to either be in the coalition or to not be in
the coalition. 
Using generating functions
\cite{Bilbao00generatingfunctions,OnlineComputation} we represent the
votes that player $A$ casts (i.e., 0 or 4) by the polynomial:
\begin{equation}
G_A 
= \frac{a^0x^0}{2} + \frac{a^4x^4}{2}
= \frac{1}{2}     + \frac{a^4x^4}{2}
\label{eq:1}
\end{equation}
Each term of this polynomial has the form $\omega a^n x^n$ where $n$
represents the number of votes cast (e.g., $a^4x^4$ means that $A$
casts 4 votes) and $\omega$ (e.g., $\frac{1}{2}$ for each term here)
represents the probability of casting that many votes for a coalition.
Note that the probabilities sum to one: $\Eval{G_A}_{a=x=1}=1$.
While previous authors used generating functions to determine voting
power, they did not include the $\omega$ factor -- they were counting
the number of coalitions, not determining the probability of each.
In this paper we call a generating function of this type a ``voting
structure''.

Similarly, the votes cast by players $\{B, C, D\}$ can be represented as
\begin{equation}
G_B 
= \frac{1}{2}
+ \frac{b^3x^3}{2}.
\qquad
G_C
= \frac{1}{2}
+ \frac{c^2x^2}{2}.
\qquad
G_D 
= \frac{1}{2}
+ \frac{dx}{2}.
\label{eq:2}
\end{equation}
The letters $\{a,b,c,d\}$ are used in order to understand the upcoming
intermediate computations; later all these variables will be given the
numerical value one.
Multiplying all four generating functions together yields
\begin{equation*}
\begin{aligned}
G_A G_B G_C G_D = \frac{1}{16}
& \left[\bigstrut\right.
  (a^4 b^3 c^2 d) x^{10} 
+ (a^4 b^3 c^2) x^9 
+ (a^4 b^3 d) x^8 
+ (a^4 b^3 + a^4 c^2 d) x^7
\\
& + (a^4 c^2 + b^3 c^2 d) x^6 
  + (a^4 d + b^3 c^2) x^5 
  + (a^4 + b^3 d) x^4
\\
& + (b^3 + c^2 d) x^3 + (c^2) x^2 + (d)x + 1 
\left.\bigstrut\right]
\\
\end{aligned}
\end{equation*}
Each term in this expression represents a coalition: the power of $x$
indicates the total votes in that coalition; the letters $\{a,b,c,d\}$
indicate the coalition composition; and the numerical coefficient
($\frac{1}{16}$ for each term) is the probability of that coalition.
For example, the $x^7$ terms shows that there are two 7 vote
coalitions: $\{A, B\}$ and $\{A, C, D\}$; each has probability
$\frac{1}{16}$ of occurring. 
Similarly there are two 6 vote coalitions: $\{A, C\}$ and $\{B, C,
D\}$; each also has a probability $\frac{1}{16}$ of occurring.

Let's focus on Player $A$. 
While all coalitions with 6 or more votes is a winning coalition, they
are not necessarily coalitions that $A$ made winning. 
For example, if the $\{A, B, C, D\}$ coalition (with 10 votes) were to
lose player $A$ then it would still have 6 votes and would still be a
winning coalition.
To identify the coalitions that $A$ can make winning, we need to start
with coalitions not involving $A$ that are not winning, add player
$A$'s votes to them, and see which ones are then winning.

To find the non-winning coalitions not involving player $A$ multiply
the generating functions for just the players $\{B, C, D\}$:
\begin{equation*}
G_B G_C G_D =
\frac{1}{8}
\surbrack{ \bigstrut
  (b^3 c^2 d) x^6 + (b^3 c^2) x^5 + (b^3 d) x^4 + (b^3 + c^2 d) x^3 
  + (c^2) x^2 + (d) x + 1
}
\end{equation*}
The coalitions that have a power $x^k$ with $k \le q-1$ are the
coalitions that are not winning.
Introduce the following notation

\NotationDefinition{
\textbf{Definition}: 
For the polynomial 
$\ds Z(x) =\sum_i\delta_ix^i$
define
$\ds \surbrace{\bigstrut Z(x)}_{\alpha}^{\beta} = 
\sum_{\alpha\le k\le \beta} \delta_kx^k$. 
\\
This extracts a set of consecutive terms in a polynomial.
}

\noindent
so that the non-winning coalitions without $A$ are:
\begin{equation}
  \Extract{ G_B G_C G_D}_0^{q-1}
= \Extract{ G_B G_C G_D}_0^{5}
= \frac{1}{8}
\surbrack{\bigstrut
  \surround{b^3 c^2    } x^5 
+ \surround{b^3 d      } x^4 
+ \surround{b^3 + c^2 d} x^3 
+ \surround{c^2        } x^2 
+ (d)x + 1
}
\label{eq:5}
\end{equation}

To determine which coalitions $A$ can make winning, multiply 
Equation (\ref{eq:5}) by $G_A$ and extract the winning coalitions,
these are the $x^k$ terms with $k \ge q = 6$:
\begin{equation}
\begin{aligned}
\Extract{ G_A \Extract{G_B G_C G_D}_0^{q-1}}_q^{\infty}
&= 
\Extract{ G_A \Extract{G_B G_C G_D}_0^{5}}_6^{\infty}
\\
&=
\frac{1}{16}
\surbrack{\bigstrut
  \surround{a^4 b^3 c^2}x^9
+ \surround{a^4 b^3 d  }x^8
+ \surround{a^4 b^3 + a^4c^2d }x^7
+ \surround{a^4 c^2 }x^6
}
\\
\end{aligned}
\label{eq:6}
\end{equation}
This shows 5 coalitions that $A$ has made winning; the first two are
$\{A, B, C\}$ and $\{A, B, D\}$.
The probability of these winning coalitions involving $A$ is the
numerical coefficient of each coalition.  

While the variables $\{a, b, c, d\}$ in Equations (\ref{eq:1}) and
(\ref{eq:2}) are useful for identifying coalitions, they are not
needed in the following.
%
%% Replacing $\{a, b, c, d\}$ with the value one makes the generating
%% functions in Equations (\ref{eq:1}) and (\ref{eq:2}) even clearer:
%% each shows a probability of $\frac{1}{2}$ for casting no votes or all
%% the votes.
%
Replacing $\{a, b, c, d\}$ with the value one in Equation (\ref{eq:6})
results in
\begin{equation*}
\Eval{
\Extract{ G_A
\Extract{ G_B G_C G_D }_0^{q-1} }_{q}^{\infty}
}_{a=b=c=d==1}
=
\frac{1}{16}
\surround{x^9+x^8+2x^7+x^6}
\end{equation*}
That is, among the coalitions that $A$ made winning there are: 2 with 7
votes and 1 with each of 6, 8, or 9 votes.
Summing the above numerical coefficients (i.e., setting $x=1$)
determines the probability that $A$ has made any coalition winning:
\begin{equation}
\begin{aligned}
\Prob_A 
&\equiv \text{Probability[Player $A$ has made a coalition winning]} \\
&= 
\Eval{\Eval{\Extract{
G_A
\Extract{G_B G_C G_D}_0^{q-1}}_q^{\infty}
}_{a=b=c=d=1}
}_{x=1}
=
\Eval{
\frac{1}{16}
\surround{x^9+x^8+2x^7+x^6}
}_{x=1}
=\frac{5}{16}
\\
\end{aligned}
\label{eq:8}
\end{equation}
This can be interpreted as follows: if one of the 16 possible
coalitions not involving $A$ were selected (uniformly) at random then
$\kfrac{5}{16}^{\text{th}}$ of the time that coalition is one for
which $A$ is marginal.

Similarly, by focusing on each of the other players one at a time, we
can compute\footnote{The intermediate computation is:
\begin{equation*}
\begin{aligned}
\Extract{ G_B \Extract{G_A G_C G_D}_0^{q-1}}_q^{\infty}
&= \frac{1}{16}
\surbrack{ \bigstrut
  \surround{b^3d} x^8 + \surround{a^4b^3}x^7 + \surround{b^3c^2d}x^6}
\\
\Extract{ G_C \Extract{G_A G_B G_D}_0^{q-1}}_q^{\infty}
&= \frac{1}{16}
\surbrack{ \bigstrut
  \surround{a^4c^2d}x^7 + \surround{a^4c^2+b^3c^2d}x^6}
\\
\Extract{ G_D \Extract{G_A G_B G_C}_0^{q-1}}_q^{\infty}
&= \frac{1}{16}
\surbrack{ \bigstrut
  \surround{b^3c^2d}x^6}
\\
\end{aligned}
\end{equation*}
}
(the subscript ``$V$'' is used at mean ``when $a = b = c = d = x = 1$''):
\begin{equation*}
\begin{aligned}
\Prob_B
&= \Eval{ \Extract{ G_B \Extract{G_A G_C G_D}_0^{q-1}}_q^{\infty} }_V
 = \frac{3}{16} \\
\Prob_C
&= \Eval{ \Extract{ G_C \Extract{G_A G_B G_D}_0^{q-1}}_q^{\infty} }_V
 = \frac{3}{16} \\
\Prob_D
&= \Eval{ \Extract{ G_D \Extract{G_A G_B G_C}_0^{q-1}}_q^{\infty} }_V
 = \frac{1}{16} \\
\end{aligned}
\end{equation*}
Computing the relative weights of these probabilities we recover the
Banzhaf powers found earlier:
\begin{equation}
\begin{aligned}
\beta(A) 
&= \frac{ \Prob_A }{ \Prob_A + \Prob_B + \Prob_C + \Prob_D }
 = \frac{ \kfrac{5}{16} }{ \kfrac{12}{16} }
 = \frac{5}{12}
\\
\beta(B) 
&= \frac{ \Prob_B }{ \Prob_A + \Prob_B + \Prob_C + \Prob_D }
 = \frac{ \kfrac{3}{16} }{ \kfrac{12}{16} }
 = \frac{1}{4}
 = \beta(C)
\\
\beta(D) 
&= \frac{ \Prob_D }{ \Prob_A + \Prob_B + \Prob_C + \Prob_D }
 = \frac{ \kfrac{1}{16} }{ \kfrac{12}{16} }
 = \frac{1}{12}
\\
\end{aligned}
\label{eq:11}
\end{equation}
Careful inspection reveals that the probabilistic computation in this
section is identical to the enumerative computation; just expressed
differently. 
The probabilities found here
$\surround{\frac{5}{16},\frac{3}{16},\frac{3}{16},\frac{1}{16}}$ are
proportional to the counts $(10,6,6,2)$ found earlier, so the voting
powers are the same.

%------------------------------------------------------
\subsection{Banzhaf Power via Influence Polynomials}
%------------------------------------------------------
%
%% In this section retain the computation in the last section, but
%% express it in a new notation. 
%% %
%% This new notation is needed to incorporate probabilistic and partial
%% voting, which are considered in the next section.
%% 
Rewrite the computation appearing in Equation (\ref{eq:8}) as
\begin{equation*}
\Prob_A
= \Eval{\Extract{
G_A
\underbrace{{\Extract{G_B G_C G_D}_0^{q-1}}}_{R(x)}
}_q^{\infty}
}_{x=1}
=
\Eval{
\Extract{
G_A
\,
R(x)
}_q^{\infty}
}_{x=1}
\end{equation*}
where $R(x)=\sum_{j=0}^{q-1} r_j x^j$ is a polynomial of degree no more than $q-1$.
(We assume now that the $\{a, b, c, d\}$ terms all have the value one.)
The constant part of $G_A$ cannot contribute to raising an exponent of
$x$ to change a non-winning coalition into a winning coalition, as
needed for the $\Extract{\cdot}_q^{\infty}$ computation, so it can be
neglected and $\Prob_A$ can be written as:
\begin{equation}
\begin{aligned}
\Prob_A 
&= \Eval{ \Extract{
\text{(Non-constant part of $G_A$)}\  R(x)
}_q^{\infty}
}_{x=1} \\
&= \Eval{ \Extract{
\surround{\frac{1}{2}x^4}
\  R(x)
}_q^{\infty}
}_{x=1} \\
&= \Eval{ \Extract{
\surround{\frac{1}{2}x^4}
\  
\surround{\sum_{j=0}^{q-1} r_j x^j}
}_q^{\infty}
}_{x=1} \\
&= 
\sum_{j=q-4}^{q-1} \frac{1}{2}r_j \\
&=
\underbrace{
\surround{\sum_{j=q-4}^{q-1} \frac{1}{2}x^j}
}_{I_A(x)}
\otimes
\surround{\sum_{j=0}^{q-1}r_jx^j}
\\
&= I_A (x)  \otimes  R(x) 
\\
\end{aligned}
\label{eq:14}
\end{equation}
where we have defined the \textit{Influence Polynomial} for $A$,
$I_A(x)$ with degree $q-1$, and we have introduced the following
notation:

\NotationDefinition{
\textbf{Definition}: For two polynomials 
  $\ds R(x) =\sum_jr_jx^j$ and
  $\ds S(x) =\sum_js_jx^j$ 
  define the sum of product coefficients to be $\ds R(x) \otimes S(x)
  = \sum_j r_js_j$.
  That is, the coefficients of common powers are multiplied together
  and then added.  }

The representation in Equation (\ref{eq:14}) is exactly equivalent to
the expression in Equation (\ref{eq:8}).
Similarly
\begin{equation}
\begin{aligned}[3]
&\Prob_B = I_B(x) \otimes 
\surbrace{\bigstrut G_A G_C G_D }_{0}^{q-1}
& \quad
&I_B(x) = \sum_{j=q-3}^{q-1}\frac{1}{2}x^j
\\
&\Prob_C = I_C(x) \otimes 
\surbrace{\bigstrut G_A G_B G_D }_{0}^{q-1}
& \quad
&I_C(x) = \sum_{j=q-2}^{q-1}\frac{1}{2}x^j
\\
&\Prob_D = I_D(x) \otimes 
\surbrace{\bigstrut G_A G_B G_C }_{0}^{q-1}
& \quad
&I_D(x) = \sum_{j=q-1}^{q-1}\frac{1}{2}x^j
\\
\end{aligned}
\label{eq:15}
\end{equation}

This section has used influence polynomials to compute the voting
probabilities for the simplest voting structure, when a voter is
equally likely to cast all or none of their votes (``random voting'').
The paradigm of using influence polynomials also works for votes
distributed partially or non-uniformly.
The next section shows how to compute the influence polynomial in
these cases.

%------------------------------------------------------
\section{Non-Uniform Probabilities}
%------------------------------------------------------
%
The generating function in Equation (\ref{eq:1}) represents the votes
that player $A$ can cast for a coalition and represents two equally
likely situations, that ``none'' or ``all'' of the available votes
were cast.
In more complex situations, weighted generating functions can capture
how players distribute their votes in ways that are not all or nothing
and to vote with non-uniform probabilities.
For example, we might choose
\begin{equation}
G_A
= \tfrac{1}{10}a^0x^0
+ \tfrac{4}{10}a^2x^2
+ \tfrac{3}{10}a^3x^3
+ \tfrac{2}{10}a^4x^4
\label{eq:16}
\end{equation}
which we interpret as follows:
Player $A$ will contribute 
0 votes to a coalition $\kfrac{1}{10}$ of the time, 
2 votes                $\kfrac{4}{10}$ of the time, 
3 votes                $\kfrac{3}{10}$ of the time, 
and
4 votes                $\kfrac{2}{10}$ of the time.

Now we must interpret what it means for a player to be “marginal” when
that player can exercise non-uniform and partial voting. 
It is no longer adequate to merely multiply the vote structures (i.e.,
generating functions) together as in Equation (\ref{eq:8}), as we now
indicate.
Imagine that player $A$ has the voting structure $G_A=x^4$; that is,
they give all 4 of their votes to \textit{every} coalition.
Blindly using Equation (\ref{eq:8}) would give $\Prob_A =\frac{5}{8}$.
This is a larger value than what was obtained in the random voting
model, and must be \textit{wrong}. 
If player $A$ \textit{always} give 4 votes to \textit{every}
coalition, then we claim that player $A$ has \textit{no} power.
This is because player $A$ has lost the ability to influence any
coalition; the other players always know what player $A$ will do, in
any circumstance.
Think of this in a political context: if a politician has already
decided to vote for (or against) a piece of legislation then they
cannot influence that legislation.
The framers of the legislation will only modify the legislation to
influence undecided voters.

In general, if a player always casts all, or none, of their votes then
that player cannot ever be marginal.
Stated differently, whenever a player cannot influence others by
having the ability to change the winning-ness of coalitions, then that
player has no power.

Let's work through an example.
Assume, as usual, that $q$ votes are needed for a coalition to be
winning.  
Suppose that a coalition not including player $A$ already has
$Z$ votes with $Z < q$ and that 
player $A$ has the vote structure in Equation (\ref{eq:16}).
Then there is a probability that each coalition without player $A$
will become, after player $A$ votes, winning ($v$) or losing ($1-v$).

Consider, for example, what this means when $v=99$\%. 
While player $A$ is nearly always giving enough votes to make the
coalition winning, the other players know that only 1\% of time will
player $A$ keep the coalition from being winning.
Hence, player $A$ will get little attention from the other players --
there is little of player $A$'s behavior that can be influenced.
Now consider instead what $v=60$\% means; more than half the time
player $A$ gives enough votes for the coalition to be winning but a
large fraction of the time (40\%) player $A$ is not giving enough
votes for a coalition to be winning.
In this case player $A$ is much more influential in determining
whether or not a coalition is going to be winning. 

We define player $A$'s ability to be marginal to be equal to the
percentage of votes that are ``in play'', the minimum of $v$ and
$1-v$; define $\gamma=\min(v, 1-v)$.
When $v=99$\% then there is only $\gamma=1$\% that is ``in play'' and
player~$A$'s influence is small; when $v=60$\% then $\gamma =40$\% and
player~$A$'s votes need to be negotiated by the other players --
player~$A$ is more of a ``swing voter'' in this case.

With this thinking the Influence Polynomial for any vote structure is
determined as follows:
\begin{enumerate} \smallspacing

\item Assume the vote structure for a player is: $\ds
  G=\sum_{j=0}^{q-1}g_j x^j$ \quad where some $\{g_j\}$ may be zero

\item Define the partial sums: $\ds v_Z =\sum_{j=q-Z}^{q-1} g_j$ and
  $\gamma_Z=\min (v_Z, 1-v_Z )$ for $Z=1,2,\dots,q-1$

\item Then the Influence Polynomial for that player is $\ds
  I(x)=\sum_{Z=1}^{q-1}\gamma_Z x^Z$

\end{enumerate}
This definition is consistent with the evaluations given earlier, for
``random voting'', as shown in the next section.
Table~\ref{tab:1} shows the Influence Polynomial computations for the
vote structure in Equation (\ref{eq:16}); the result is
\begin{equation*}
I_A(x) =
0x^1
+\tfrac{2}{10} x^2
+\tfrac{5}{10} x^3
+\tfrac{1}{10} x^4
+\tfrac{1}{10} x^5
\end{equation*}

\renewcommand\ttemp[1]{ \frac{#1}{10} }
\renewcommand\temp[5]{ 
\bigstrut
$#1$ & 
$#2$ & 
$#3$ & 
$#4$ & 
$#5$ \\
\hline
}

\begin{table}[t]
\begin{center}
\begin{tabular}{|c|l|c|c|c|}
\hline
\ZboxStrut{1.2in}{%
Number of votes coalition has without player $A$}
&
\ZboxStrut{1.3in}{%
Probability of  coalition  winning with $A$'s  votes: $v_Z$}
&
\ZboxStrut{1.3in}{%
Probability of coalition \textit{not} winning with $A$'s votes: 
$(1-v_Z)$}
&
\ZboxStrut{1.4in}{%
Fraction of $A$'s votes that are ``in play'':
 $\gamma_Z = \min(v_Z , 1-v_Z )$}
&
$x^Z$
\\
\hline
\temp{Z=1}{ 0}{ 1}{ 0}{ x^1}
\temp{Z=2}{\frac{2}{10}}{\frac{8}{10}}{\frac{2}{10}}{x^2}
\temp{Z=3}
{\ttemp{5}=\ttemp{2}+\ttemp{3}}
{\ttemp{5}}
{\ttemp{5}}
{x^3}
\temp{Z=4}
{\ttemp{9}=\ttemp{2}+\ttemp{3}+\ttemp{4}}
{\ttemp{1}}
{\ttemp{1}}
{x^4}
\temp{Z=5}
{\ttemp{9}=\ttemp{2}+\ttemp{3}+\ttemp{4}}
{\ttemp{1}}
{\ttemp{1}}
{x^5}
\end{tabular}
\end{center}
\caption{A coalition without player $A$ has $Z$ votes and a winning coalition needs $q=6$ votes; player $A$ votes
using the vote structure in Equation (\ref{eq:16}).}
\label{tab:1}
\end{table}

Using the Influence Polynomial $I_A(x)$ we define the
\textit{Influence of $A$}, $I(A)$, to be:
\begin{equation}
I(A)=I_A(x) \otimes
\Extract{  G_B G_C G_D }_0^{q-1}
\label{eq:influence}
\end{equation}
which is a generalization of the probability defined in Equations
(\ref{eq:14}) and (\ref{eq:15}). 
This becomes the probability shown in those equations when a player is
using ``random voting''.
Once the influences have been determined for each player, they are
normalized as in Equation (\ref{eq:11}) to determine what we define to
be the Generalized Banzhaf power; for player $A$ this is denoted
$\beta'(A)$.

The Generalized Banzhaf power is a generalization of the Banzhaf power
that accounts for arbitrary voting structures.
For random voting, the Generalized Banzhaf power is the Banzhaf power.

%------------------------------------------------------
\subsection{Influence Polynomial for Random Voting}
%------------------------------------------------------
%
The Influence Polynomials as defined algorithmically in the last
section is consistent with the values given in Equation (\ref{eq:15}),
as we now show. 
Assume use of random voting, that is:
\begin{equation*}
G_N
=\frac{1}{2} + \frac{X^N}{2}
=\frac{1}{2} \sum_{j=0}^{q-1} \surround{\delta_{j0}+\delta_{jN}}x^j
\end{equation*}
where $\delta_{ij}$ is the usual Kronecker delta and $N\le q-1$.
Using the procedure for determining the Influence Polynomial in the
last section (recall $Z\le q-1$), we compute
\begin{equation*}
\begin{aligned}
v_Z 
&=\sum_{j=q-Z}^{q-1} g_j
= \frac{1}{2} \sum_{j=q-Z}^{q-1} \surround{\delta_{j0}+\delta_{jN}}
=\begin{cases}
  \frac{1}{2} & Z\ge q-N \\
  0           & Z<q-N \\
 \end{cases}
\\
\gamma_Z
&=\min(v_Z, 1-v_Z ) 
=\begin{cases}
  \frac{1}{2} & Z\ge q-N \\
  0           & \text{otherwise} \\
 \end{cases}
\\
I(x)
&=\sum_{Z=1}^{q-1}\gamma_Z x^Z
=\frac{1}{2}\sum_{Z=q-N}^{q-1} x^Z
=\frac{1}{2}\surround{x^{q-N}+x^{q-N+1}+\cdots+x^{q-1}}
\\
\end{aligned}
\end{equation*}
If, for example, $q=6$ and $N=3$ then 
$\ds I(x)=\frac{1}{2}\surround{x^3+x^4+x^5}$
as shown in Equation (\ref{eq:15}) for player~$B$.

%------------------------------------------------------
\subsection{Example: [6;4,3,2,1] game with one player having non-uniform votes}
%------------------------------------------------------
We assume the voting structures appearing in Equations (\ref{eq:2}) and (\ref{eq:16})
\begin{equation*}
\begin{aligned}
G_A &= \tfrac{1}{10} + \tfrac{4}{10}x^2 + \tfrac{3}{10}x^3 + \tfrac{2}{10}x^4, \\
G_B &= \tfrac{1}{2} + \tfrac{1}{2}x^3, \qquad
G_C  = \tfrac{1}{2} + \tfrac{1}{2}x^2, \qquad
G_D  = \tfrac{1}{2} + \tfrac{1}{2}x
\\
\end{aligned}
\end{equation*}
for which the Influence Polynomials have been determined to be:
\begin{equation*}
\begin{aligned}
I_A(x) &= \tfrac{2}{10}x^2 + \tfrac{5}{10}x^3 + \tfrac{1}{10}x^4 + \tfrac{1}{10}x^5 \\
I_B(x) &= \tfrac{1}{2}\surround{ x^3+x^4+x^5} \\
I_C(x) &= \tfrac{1}{2}\surround{     x^4+x^5} \\
I_D(x) &= \tfrac{1}{2}\surround{         x^5} \\
\end{aligned}
\end{equation*}
Using Equation (\ref{eq:influence}) and its analogues we find the
influences $\{ I(A), I(B), I(C), I(D)\}$.
Normalizing the influences by their sum gives the Generalized
Banzhaf powers $\{ \beta'(A), \beta'(B), \beta'(C), \beta'(D)\}$:
\begin{equation*}
\begin{aligned}
  I(A) &= \tfrac{ 7}{40}, \qquad
& I(B) &= \tfrac{13}{40}, \qquad
& I(C) &= \tfrac{ 3}{20}, \qquad
& I(D) &= \tfrac{ 1}{10}, \qquad
\\
  \beta'(A) &= \tfrac{ 7}{30}, \qquad
& \beta'(B) &= \tfrac{13}{30}, \qquad
& \beta'(C) &= \tfrac{ 6}{30}, \qquad
& \beta'(D) &= \tfrac{ 4}{30}, \qquad
\\
\end{aligned}
\end{equation*}

%------------------------------------------------------
\subsection{Example: [6;4,3,2,1] game with one player voting parametrically}
%------------------------------------------------------
%
For the $[6; 4, 3, 2, 1]$ game suppose that players $B$, $C$, and $D$
vote as before; that is, using random voting (each is equally likely
to give no votes or all votes).
Suppose now that player~$A$ gives 0 votes with probability $1-p$ and
gives 4 votes with probability $p$; that is player~$A$ has the
parametric vote structure (with $0\le p \le 1$)
\begin{equation*}
G_A=(1-p) + px^4
\end{equation*}
For this voting structure, 
$I_A(x)=\surround{x^2 + x^3 + x^4 + x^5}\min(p, 1-p)$ and
\begin{equation}
\beta'(A) = \frac{5 \min(1-p, p)}{\Delta_4}, \qquad
\beta'(B) = \beta'(C) = \frac{1+p}{\Delta_4}, \qquad
\beta'(D) = \frac{1-p}{\Delta_4}
\label{eq:25}
\end{equation}
where $\Delta_4=3 + p + 5 \min(1-p, p)$. 
These results are shown in Figure \ref{fig:1}.
Observe that:

\begin{figure}
\parbox{\hsize}{
\hfil
  \scalebox{0.45}{\includegraphics{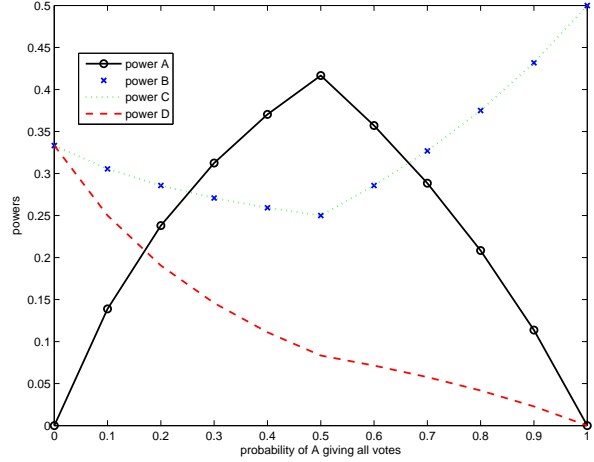}}
\hfil}
% GOOD \includegraphics{zfig_power4vary321c.eps}
% BAD \includegraphics[height=3in]{zfig_power4vary321c.eps}
%
\caption{Generalized Banzhaf powers for $[6;4,3,2,1]$ game when
  player~$A$ has vote structure $G_A=(1-p) + px^4$.}
\label{fig:1}
\end{figure}

\begin{figure}
\parbox{\hsize}{
\hfil
\scalebox{0.31}{\includegraphics{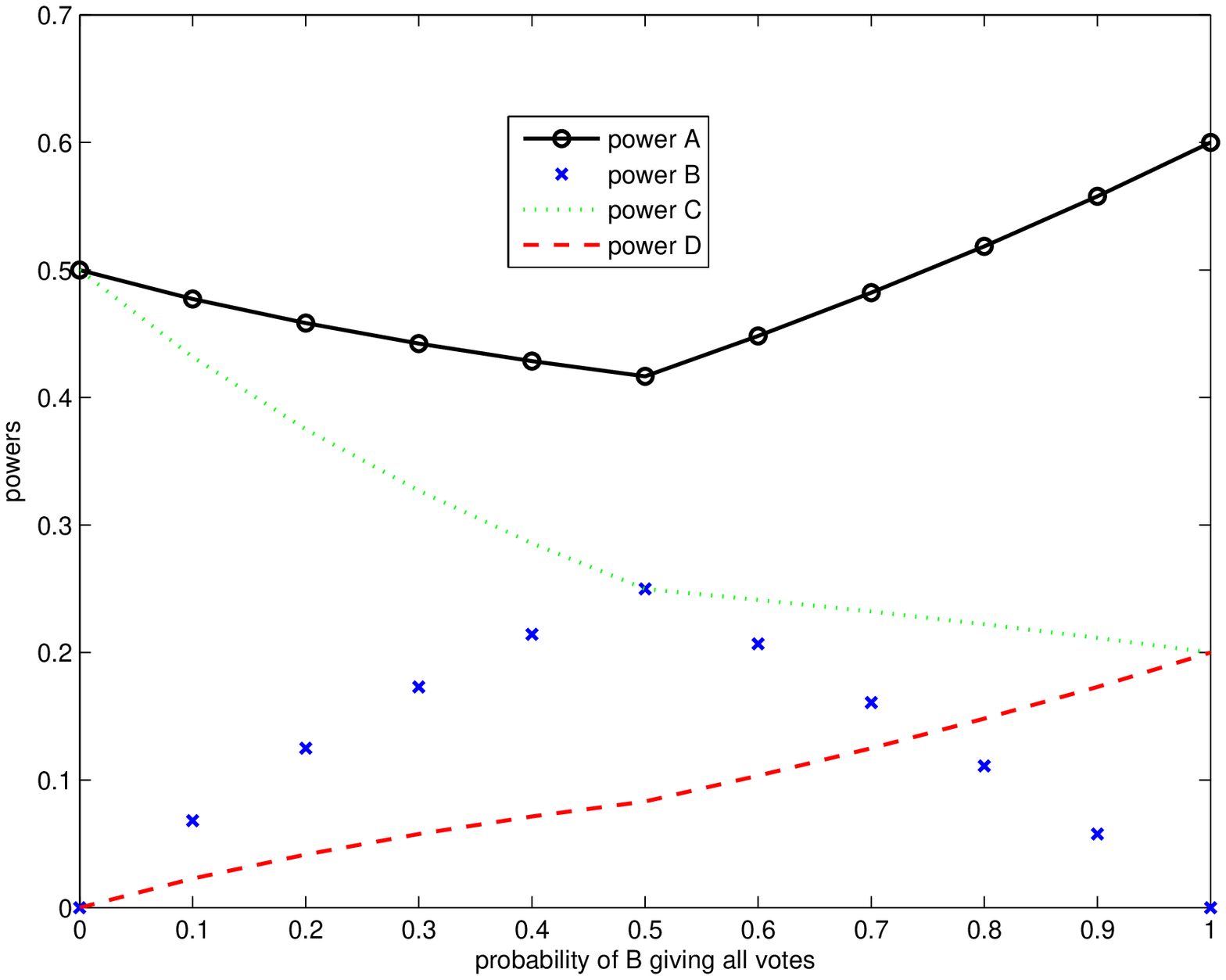}}\hfil
\scalebox{0.31}{\includegraphics{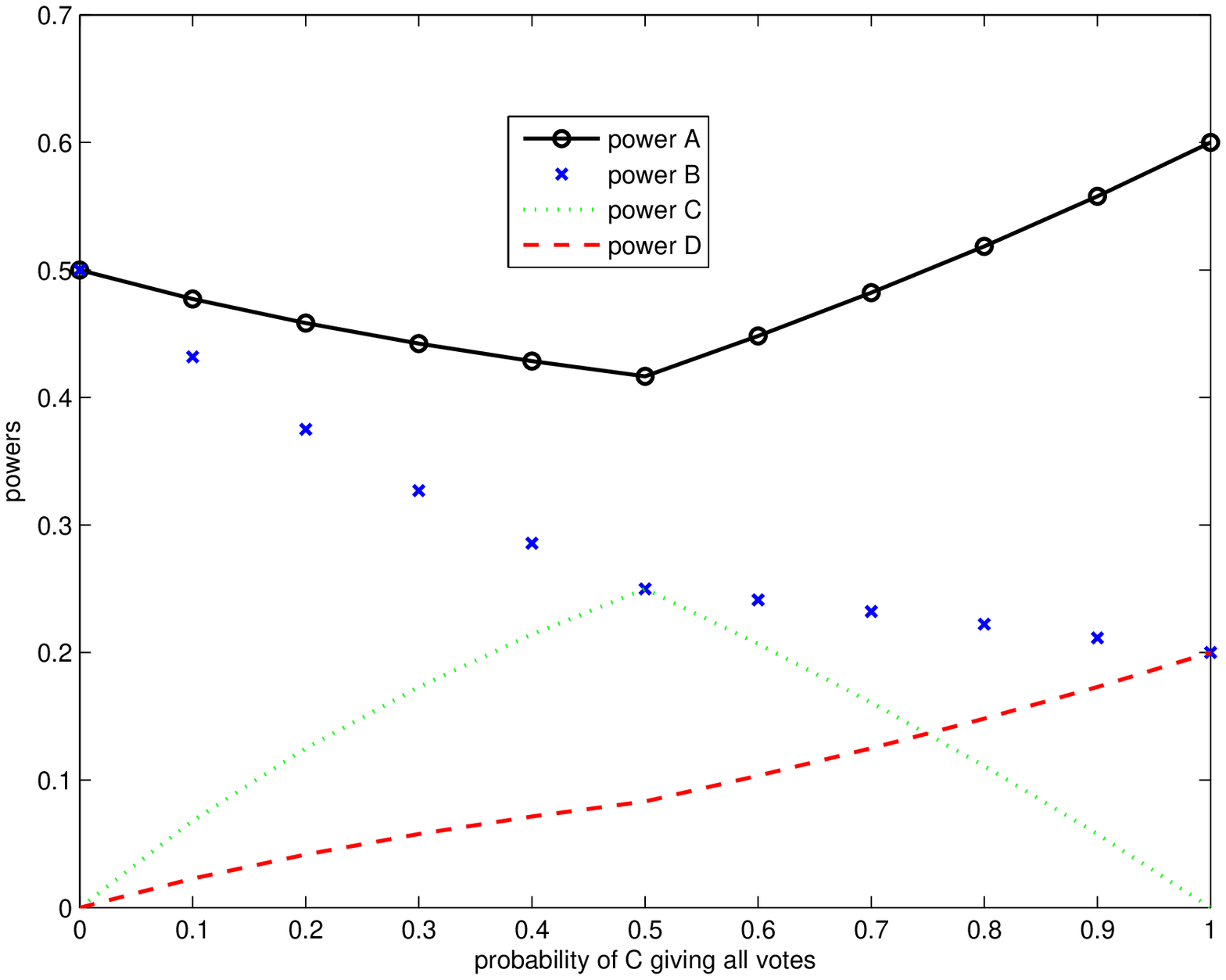}}\hfil
\scalebox{0.31}{\includegraphics{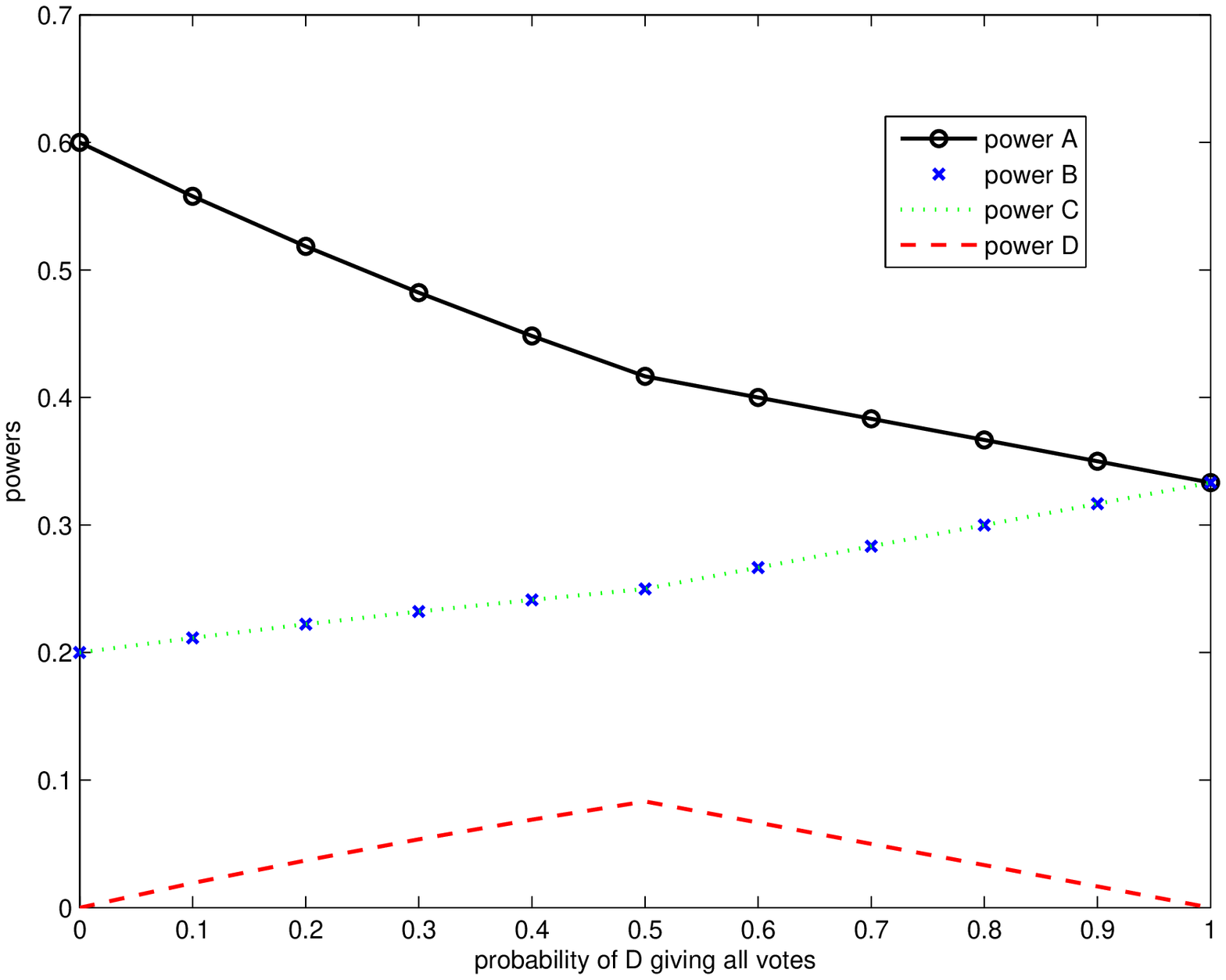}}\hfil
}
\caption{The Generalized Banzhaf powers for the voting structures in
Equations (\ref{eq:26}), (\ref{eq:27}), and (\ref{eq:28}).}
\label{fig:2}
\end{figure}

\begin{enumerate}

\item Player~$A$ has a Generalized Banzhaf power of zero when $p=0$ or
  $p=1$.
  This is expected, player~$A$ has no power when there are no votes
  ``in play''.

\item Player~$A$ has a maximal Generalized Banzhaf power when
  $p=\frac{1}{2}$ .
This is expected, this is when player A has the most votes ``in play''.

\item Players $B$ and $C$ always have the same Generalized Banzhaf
  power.

\item When $p=0$ (player~$A$ casts no votes) the game is the same as
  $[6; 3, 2, 1]$ for the players $\{B,C,D\}$.
In this case players $B$, $C$, and $D$ all have equal Generalized
Banzhaf power of~$\frac{1}{3}$, which is the same as their Banzhaf
power.

\item When $p=1$ (player~$A$ casts 4 votes) the game is the same as
  $[2; 3, 2, 1]$ for the players $\{B,C,D\}$; this is an improper
  game, but the meaning is clear.
  In this case players $B$ and $C$ have equal Generalized Banzhaf
  power of $\frac{1}{2}$ and player~$D$ has a Generalized Banzhaf
  power of zero.

\end{enumerate}
In the $[6; 4, 3, 2, 1]$ game a player other than player $A$ could vote
parametrically.
In the following three examples player $B$, $C$, or $D$ gives 0 votes
with probability $1-p$ and gives all its votes with probability $p$;
in each case the other players use random voting.
Figure \ref{fig:2} shows the results graphically.
\begin{enumerate}

\item The voting structures and Generalized Banzhaf powers when player
  $B$ votes parametrically:
\begin{equation}
\begin{aligned}
  G_A &= \tfrac{1}{2} \surround{1+x^4}, %\quad
& G_B &= (1-p)+px^3,                    %\quad
& G_C &= \tfrac{1}{2} \surround{1+x^2}, %\quad
& G_D &= \tfrac{1}{2} \surround{1+x}
\\
  \beta'(A) &= \frac{2+p           }{\Delta_3}, %\quad
& \beta'(B) &= \frac{3 \min(1-p, p)}{\Delta_3}, %\quad
& \beta'(C) &= \frac{2-p           }{\Delta_3}, %\quad
& \beta'(D) &= \frac{p             }{\Delta_3}
\\
\end{aligned}
\label{eq:26}
\end{equation}
where $\Delta_3=4+p+3\min(1-p,p)$

\item The voting structures and Generalized Banzhaf powers when player
  $C$ votes parametrically:
\begin{equation}
\begin{aligned}
  G_A &= \tfrac{1}{2} \surround{1+x^4}, %\quad
& G_B &= \tfrac{1}{2} \surround{1+x^3}, %\quad
& G_C &= (1-p)+px^2,                    %\quad
& G_D &= \tfrac{1}{2} \surround{1+x}
\\
  \beta'(A) &= \frac{2+p           }{\Delta_2}, %\quad
& \beta'(B) &= \frac{2-p           }{\Delta_2}, %\quad
& \beta'(C) &= \frac{3 \min(1-p, p)}{\Delta_2}, %\quad
& \beta'(D) &= \frac{p             }{\Delta_2}
\\
\end{aligned}
\label{eq:27}
\end{equation}
where $\Delta_2=4+p+3\min(1-p,p)$

\item The voting structures and Generalized Banzhaf powers when player
  $D$ votes parametrically:
\begin{equation}
\begin{aligned}
  G_A &= \tfrac{1}{2} \surround{1+x^4}, %\quad
& G_B &= \tfrac{1}{2} \surround{1+x^3}, %\quad
& G_C &= \tfrac{1}{2} \surround{1+x^2}, %\quad
& G_D &= (1-p)+px
\\
  \beta'(A) &= \frac{3-p}{\Delta_1}, %\quad
& \beta'(B) &= \frac{1+p}{\Delta_1}, %\quad
& \beta'(C) &= \frac{1+p}{\Delta_1}, %\quad
& \beta'(D) &= \frac{\min(1-p,p) }{\Delta_1}
\\
\end{aligned}
\label{eq:28}
\end{equation}
where $\Delta_1=5+p+\min(1-p,p)$

\end{enumerate}

\newpage
%------------------------------------------------------
\section{Teams and Leaders}
%------------------------------------------------------
%
Another generalization of traditional voting is to consider ``teams''
(or coalitions) of players that work together, although not with
complete unanimity.
For example, for the $[6; 4, 3, 2, 1]$ game assume that player~$A$
(with 4 votes) represents a team (shown as $\Ateam$) of 3 members
$\{a_1 , a_2 , a_3 \}$ with the first two members having 1 vote each and
the last member having 2 votes.

Suppose the following: 
\begin{enumerate}

\item $\Ateam$ has a leader who influences how the $\Ateam$ members
  cast their votes.
We define the leader's power to the same as their team's power.

\item The $\Ateam$ leader wants each individual $\Ateam$ member to
  cast their votes with probability $L$ and to not cast their votes
  with probability $(1-L)$.

\item Each individual $\Ateam$ member follow their leader's desire
  with probability $p$ and each member does so independently of other
  team members.

\end{enumerate}
In this case the appropriate generating function representation of
$\Ateam$'s votes is
\begin{equation}
\begin{aligned}
G_{\Ateam} 
\quad
=
\qquad\qquad
L&
\surbrack{
\underbrace{\surround{\bigstrut (1-p)+a  px  }}_{\text{member $a_1$}}
\underbrace{\surround{\bigstrut (1-p)+a  px  }}_{\text{member $a_2$}}
\underbrace{\surround{\bigstrut (1-p)+a^2px^2 }}_{\text{member $a_3$}}
}\\
+(1-L)&
\surbrack{
\underbrace{\surround{\bigstrut p+a  (1-p)x  }}_{\text{member $a_1$}}
\underbrace{\surround{\bigstrut p+a  (1-p)x  }}_{\text{member $a_2$}}
\underbrace{\surround{\bigstrut p+a^2(1-p)x^2 }}_{\text{member $a_3$}}
}
\\
\end{aligned}
\label{eq:29}
\end{equation}
The first term (with the $L$ coefficient) represents the votes cast if
the leader wishes $\Ateam$ to be part of a coalition, the second term
(with the $(1-L)$ coefficient) represents the votes cast if the leader
wishes $\Ateam$ to not be part of a coalition. 
The generating functions for each member are multiplied together, in
each sub-expression, since each team member acts independently.

As before, this generating function has $x$ exponents of $0,1,\dots,4$
representing the number of votes that $\Ateam$ can cast.
Note that the expression is correctly normalized; 
$\Eval{G_{\Ateam}}_{a=x=1}=1$ for any value of~$p$.
Table~\ref{tab:2} interprets $G_{\Ateam}$ for specific values of $L$
and ~$p$.

\renewcommand\temp[3]{%
% \bigstrut
\Zbox{0.9in}{$\ds #1$} & 
$\ds #2$ & 
\Zbox{3.5in}{{\ }\\ #3\\}  \\
\hline
}

\begin{table}[!tbh]
\begin{center}
\begin{tabular}{|l|c|l|}
\hline
 \bigstrut 
 \textbf{Parameter values}
&\textbf{Value of $G_{\Ateam}$}
&\textbf{Interpretation}
\\
\hline
\temp{p=1}
{L \surround{a^4 x^4} + (1-L)}
{All players vote exactly as their leader wishes.  
Structurally this has the form of one player voting parametrically.}
\temp{p=1, L =\frac{1}{2}}
{\frac{1}{2}+\frac{a^4x^4}{2}}
{Players vote exactly as the leader wishes and the leader is equally likely to
  support or oppose joining a coalition. $G_{\Ateam}$ is the same as 
  $G_A$ in Equation (\ref{eq:1}).}
\temp{p=1, L=1}
{a^4 x^4}
{Players vote exactly as the leader wishes and the leader
  wants to join a coalition. All 4 votes are cast.}
\temp{p=1, L=0}
{1}
{Players vote exactly as the leader wishes and the
  leader is opposed to joining a coalition. No votes are
  cast.}
\temp{p=\frac{1}{2}}
{\frac{1}{8}(ax+1)^2(a^2x^2+1)}
{Players vote randomly  and are not following their leader. 
  $G_{\Ateam}$ does not depend on $L$.}
\temp{p=0}
{L+(1-L)a^4x^4}
{Players do the exact opposite of what their leader wants.
Structually this has the form of one player voting parametrically.}
\temp{p\to 1-p$ and $L\to 1-L}
{G_{\Ateam}}
{If the leader switches their desire to join a coalition
and the players switch their likelihood of following
their leader, the result is the same.}
\end{tabular}
\end{center}
\caption{Interpretation of $G_{\Ateam}$ from Equation (\ref{eq:29}) for selected parameter values.}
\label{tab:2}
\end{table}

%------------------------------------------------------
\subsection{Teams whose members each have one vote}
\begin{figure}[!tbh]
\parbox{\hsize}{
\hfil
\scalebox{0.3}{\includegraphics{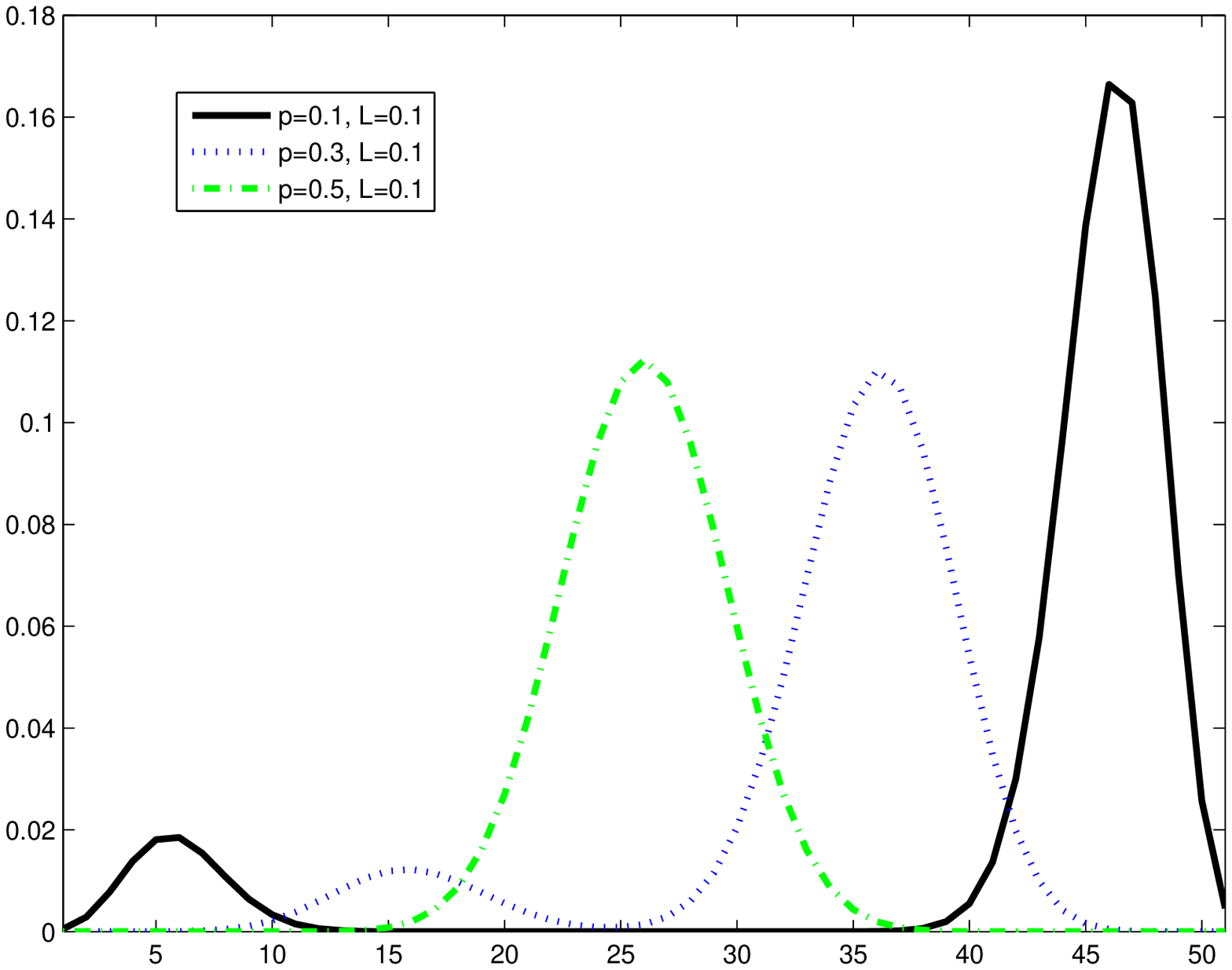}}\hfil
\scalebox{0.3}{\includegraphics{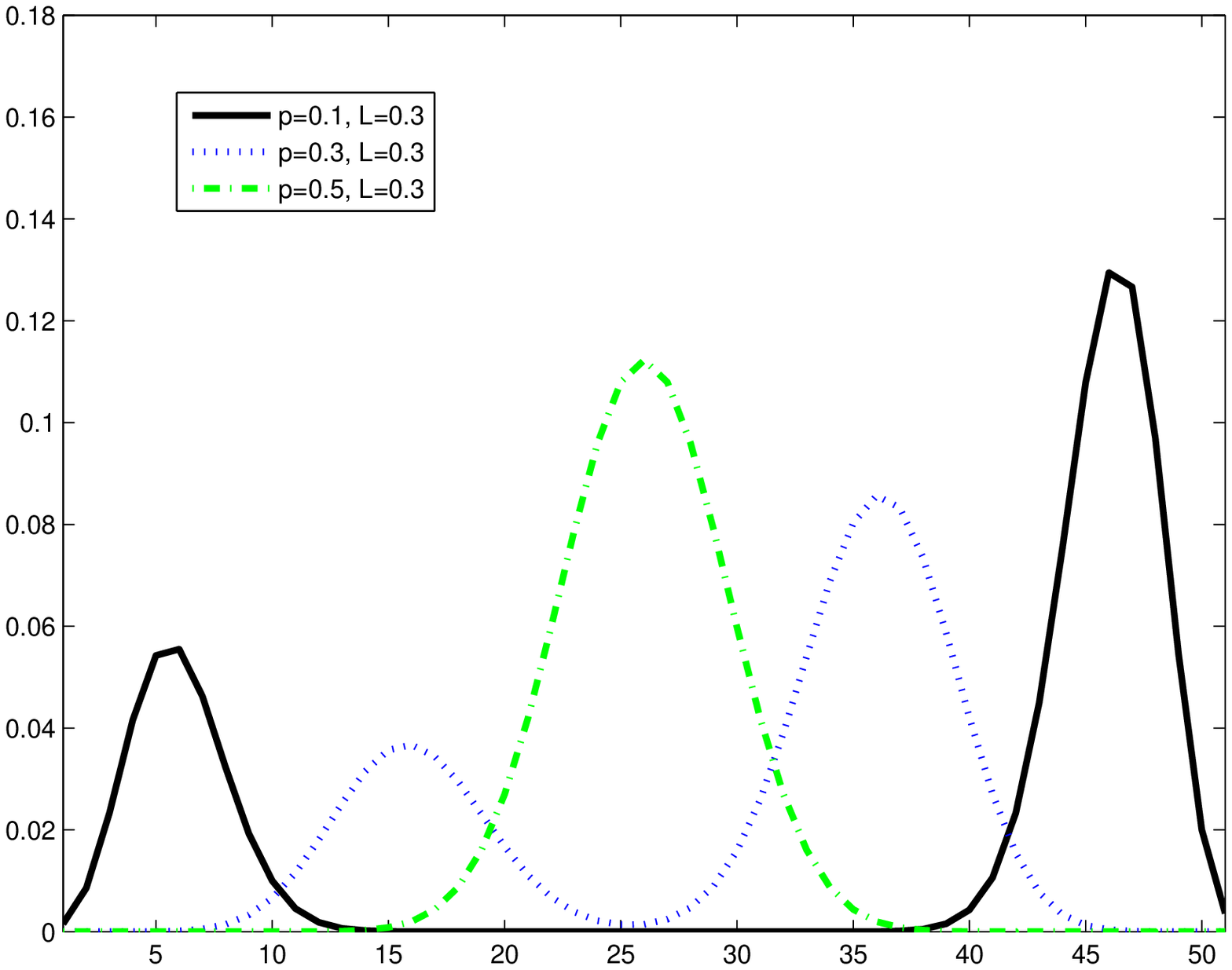}}\hfil
\scalebox{0.3}{\includegraphics{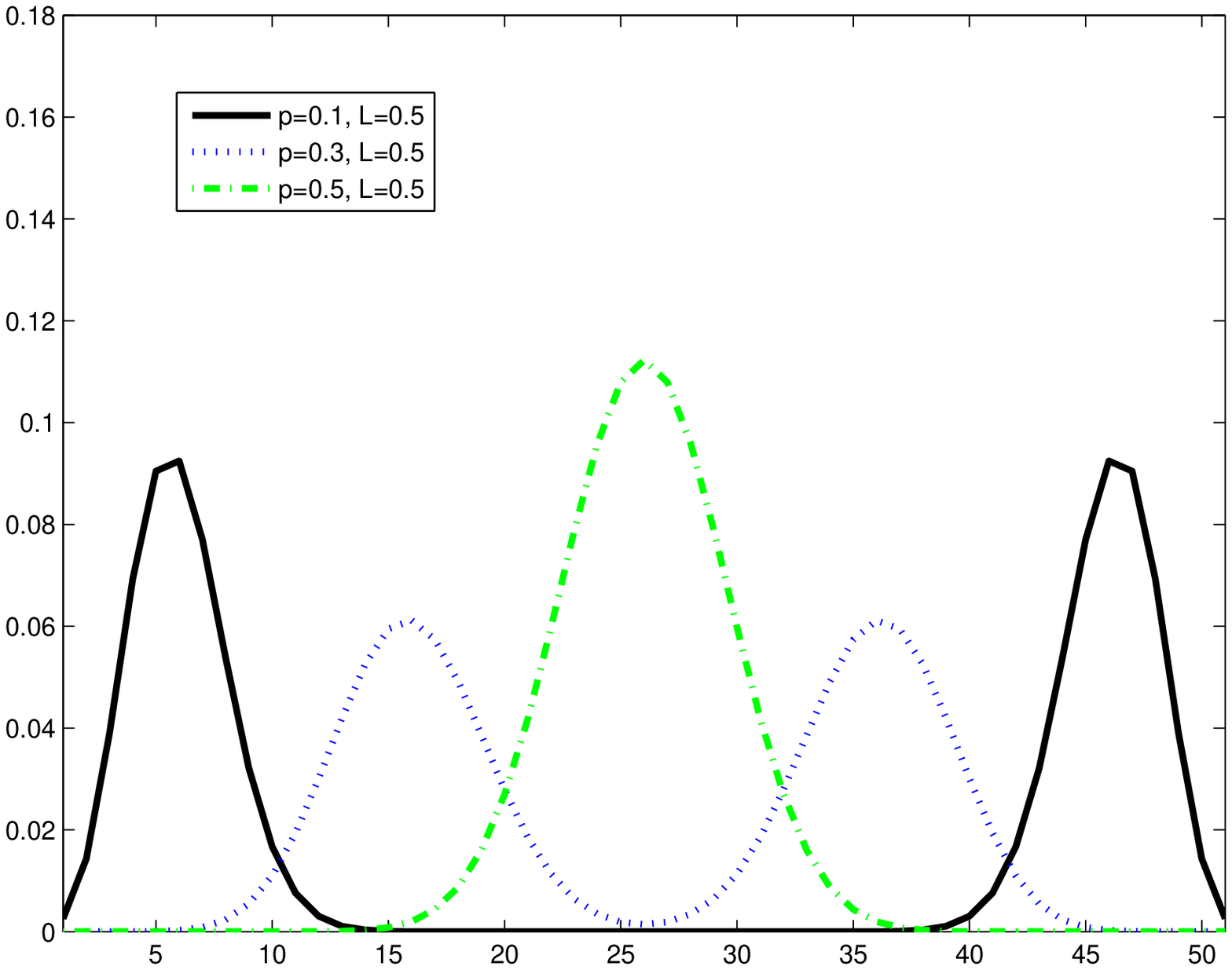}}\hfil
}
\caption{The polynomial coefficients of $G_{\text{uniform team}}$ in
  Equation (\ref{eq:30}) when $n=50$ for $p=0.1$ (black, solid line),
  $p=0.3$ (blue, dotted line), and $p=0.5$ (green, dashed line).
  Values of $L$ are $L=0.1$ (left), $L=0.3$ (middle), and 
  $L=0.5$ (right).
%
%  Left graph has $L=0.1$, middle graph has $L=0.3$, right graph has
%  $L=0.5$.
}
\label{fig:large:n:generating:function}
\end{figure}

\begin{figure}[!tbh]
\parbox{\hsize}{
\hfil
\scalebox{0.3}{\includegraphics{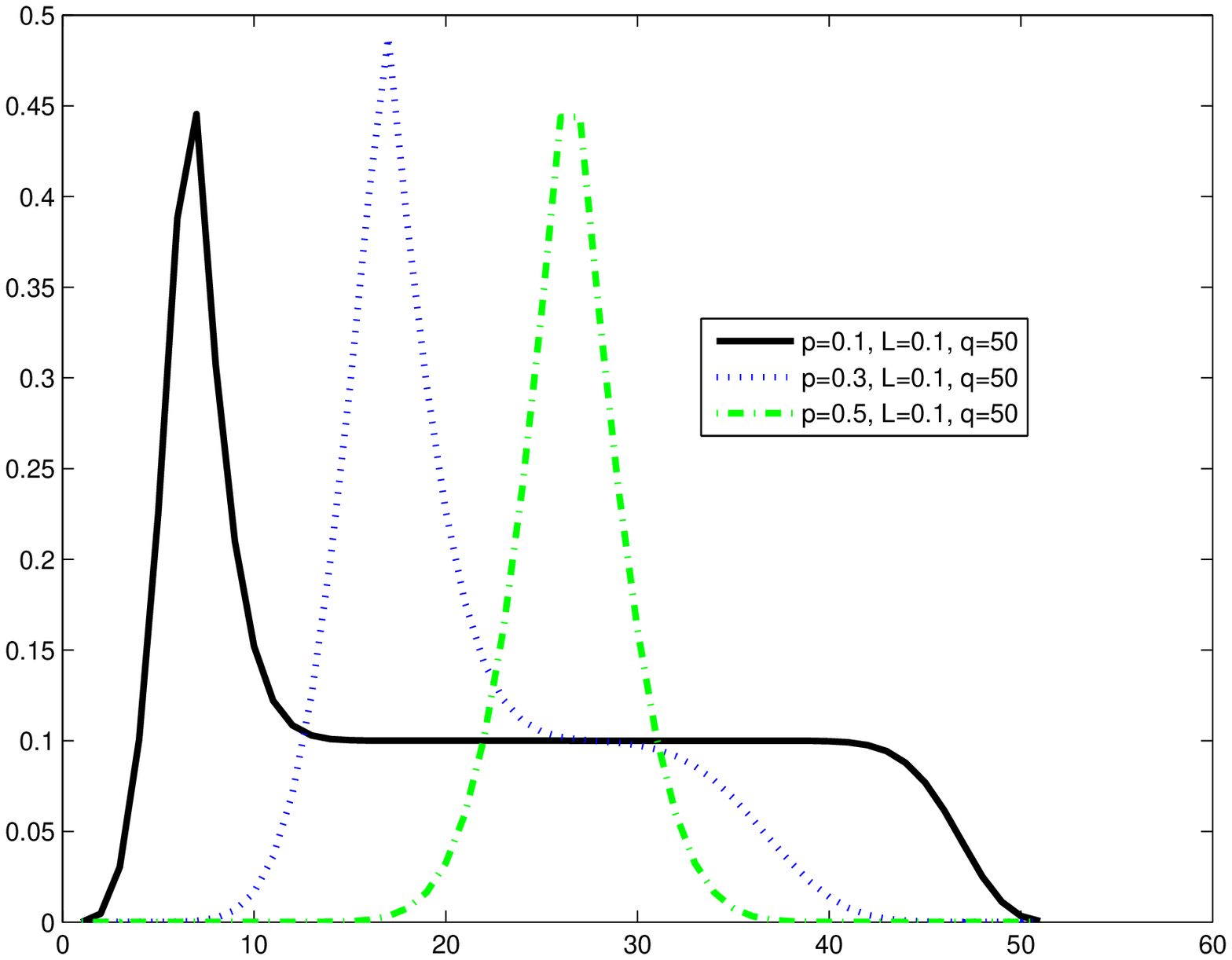}}\hfil
\scalebox{0.3}{\includegraphics{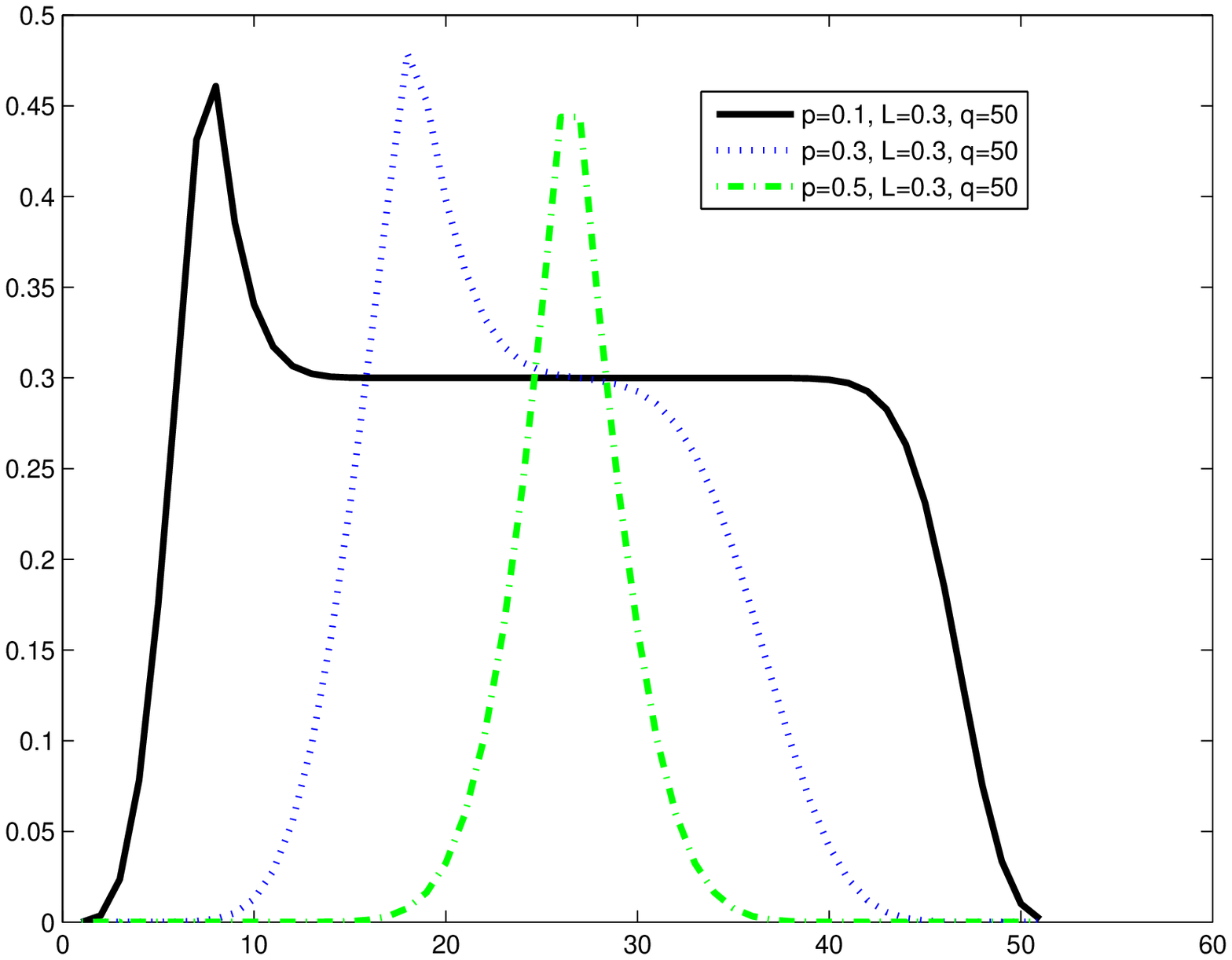}}\hfil
\scalebox{0.3}{\includegraphics{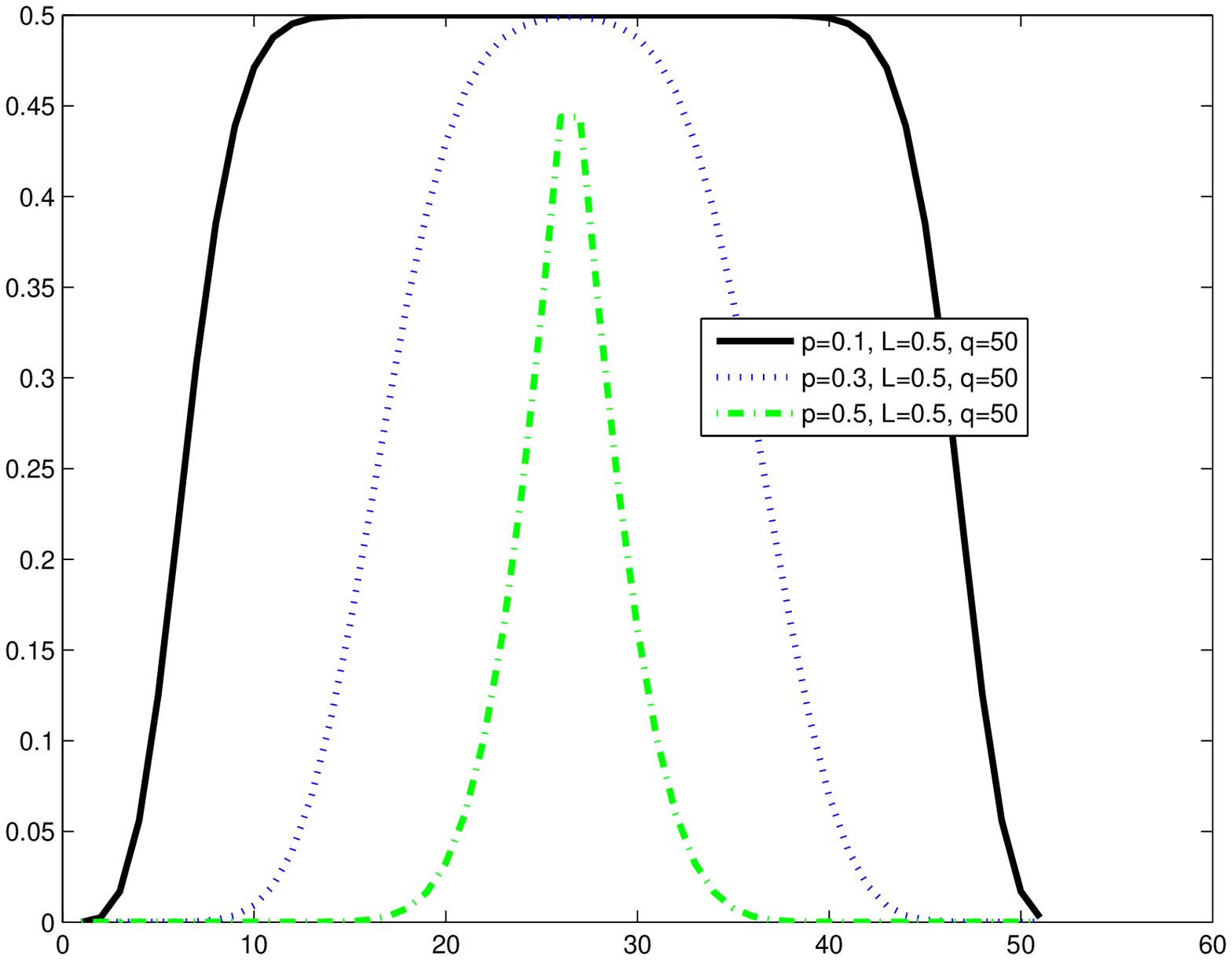}}\hfil
}
\caption{The polynomial coefficients of the Influence Polynomial
  obtained from $G_{\text{uniform team}}$ in Equation
  (\ref{eq:30}). The parameters $\{n,p,L\}$ are the same as in
  Figure~\ref{fig:large:n:generating:function}.}
\label{fig:large:n:influence:polynomial}
\end{figure}

%------------------------------------------------------
%
An important special case is a team whose members each have one vote.
For example, this could represent Congress where each Congressperson
has one vote for their team; and the teams are called Democrats,
Republicans, or Independents.
The voting structure for a team ($G_{\text{uniform team}}$) of $n$
members, where each member has a single vote is:
\begin{equation}
G_{\text{uniform team}}
=    L  \surround{\bigstrut (1-p) + px}^n
+ (1-L) \surround{\bigstrut p + (1-p)x}^n
\label{eq:30}
\end{equation}
Special cases of this are:
\begin{itemize} % \smallspacing

\item
If $p=\frac{1}{2}$
then $G_{\text{uniform team}}=\surround{\frac12+\frac12 x}^n$ independent of $L$.
\\
(This is reasonable, team members are not influenced by their leader's choice.)

\item If $p=1$ then $G_{\text{uniform team}}=Lx^n + (1-L)$
  \\
  (This is reasonable, with complete unanimity the team acts like one
  voter who distributes all the votes or none of the votes.)
\end{itemize}
Figure~\ref{fig:large:n:generating:function} shows the coefficients of $G_{\Ateam}$ when $n=50$
for various values of $L$ and~$p$.
Since $n$ is large the coefficients closely approximate either a
Gaussian (when $p=\frac12$) of the sum of two Gaussians.
Figure~\ref{fig:large:n:influence:polynomial} shows the coefficients
of the Influence Polynomials for $G_{\Ateam}$ when $n=50$ for the same
values of $L$ and~$p$.

%------------------------------------------------------
\subsection{Example: [6;4,3,2,1] game when first player is a team}
\begin{figure}
\parbox{\hsize}{
\hfil
  \scalebox{0.4}{\includegraphics{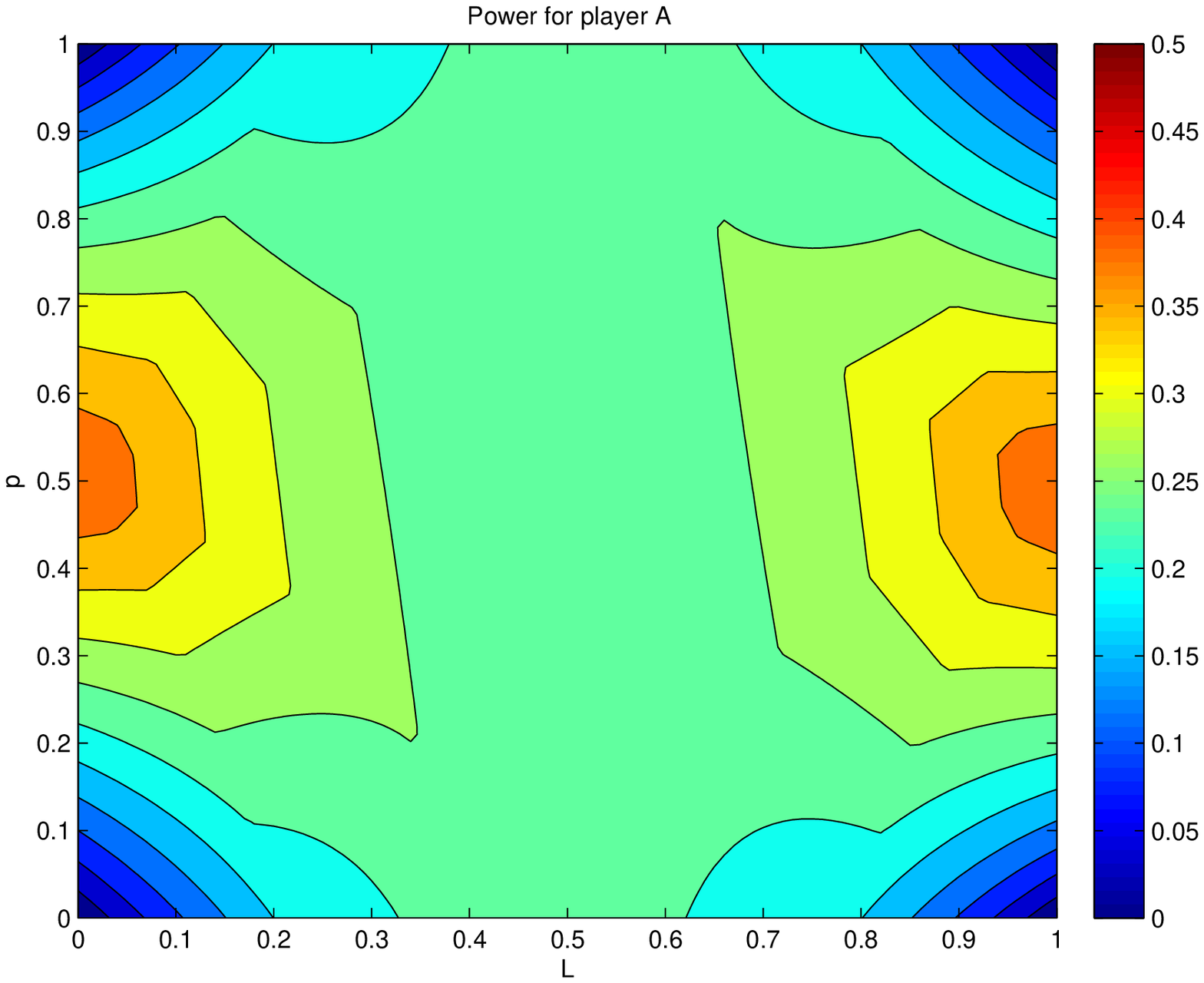}} \hfil
  \scalebox{0.4}{\includegraphics{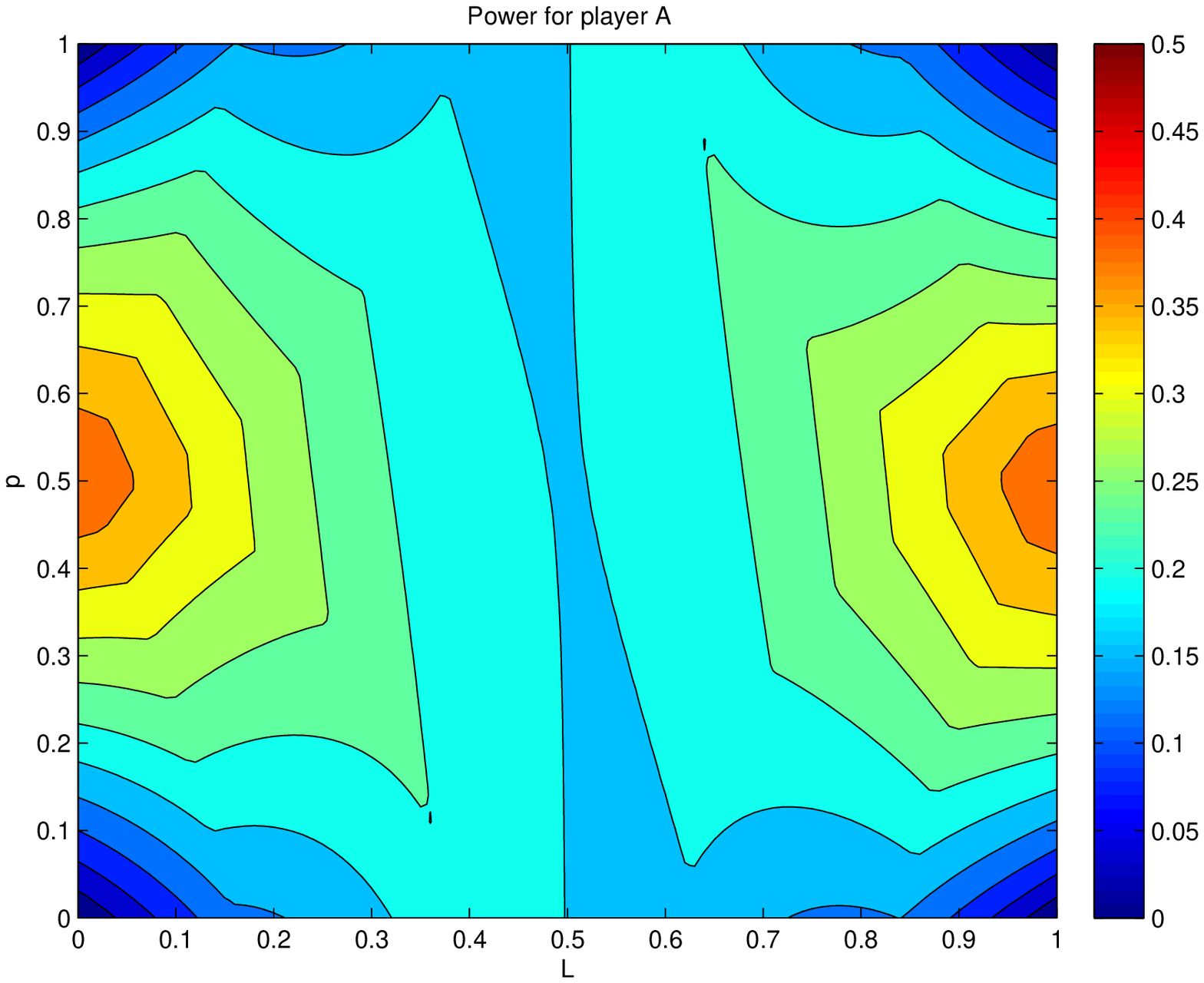}} \hfil
}
\caption{Contour plots of $\beta'(A)$ when team $A$ has 3 members
  (left, Equation (\ref{eq:31})) or 4 members (right, Equation
  (\ref{eq:32})). 
  The $L$ axis (0 to 1) is horizontal, the $p$ axis
  ($\frac12$ to 1) is vertical.  
  For all the contour plots in this paper the color scale goes from 0
  (blue) to $\frac{1}{2}$ (red).}
\label{fig:4}
\end{figure}

%------------------------------------------------------
%
Consider a voting structure where $\Ateam$ has 3 members (using
Equation (\ref{eq:29}) with individual weights of $\{2,1,1\}$) while
the other players use uniform voting
\begin{equation}
\begin{aligned}
&G_{\Ateam} = 
L
\surbrack{
\surround{\bigstrut (1-p)+ px   }^2
\surround{\bigstrut (1-p)+ px^2 }
}
+(1-L)
\surbrack{
\surround{\bigstrut p+ (1-p)x   }^2
\surround{\bigstrut p+ (1-p)x^2 }
}
\\
&
G_B = \tfrac{1}{2} + \tfrac{1}{2} x^3, \qquad
G_C = \tfrac{1}{2} + \tfrac{1}{2} x^2, \qquad
G_D = \tfrac{1}{2} + \tfrac{1}{2} x  
\\
\end{aligned}
\label{eq:31}
\end{equation}
the Generalized Banzhaf power for team $A$ is shown in Figure
\ref{fig:4} (left) as a function of $L$ and $p$.  

Now consider a voting structure where $\Ateam$ has 4 identical members
(using Equation (\ref{eq:30}), each team $A$ member has 1 vote) while
the other players use uniform voting
\begin{equation}
\begin{aligned}
&G_{\Ateam} = 
    L \surround{\bigstrut (1-p)+ p    x }^4
+(1-L)\surround{\bigstrut    p + (1-p)x }^4
\\
&
G_B = \tfrac{1}{2} + \tfrac{1}{2} x^3, \qquad
G_C = \tfrac{1}{2} + \tfrac{1}{2} x^2, \qquad
G_D = \tfrac{1}{2} + \tfrac{1}{2} x  
\\
\end{aligned}
\label{eq:32}
\end{equation}
the Generalized Banzhaf power for team $A$ is shown in Figure
\ref{fig:4} (right) as a function of $L$ and $p$.

In both of these cases:
\begin{enumerate} \smallspacing

\item The symmetry represented by $\{p,L\}\to\{1-p,1-L\}$ is apparent

\item For any value of $L$ the maximum power for each $\Ateam$ is attained
  when $p=\frac12$.

\item When $p=\frac12$ the maximum power for each $\Ateam$ is attained when
  $L=0$ or $L=1$.

\item When $p$ is zero or one and $L$ is zero or one then $\Ateam$ has
  zero power.

\end{enumerate}
In each case, $\Ateam$ has the most power when the members are least
predictable ($p=\frac12$) and the leader is decisive (either $L=0$ or
$L=1$)

%------------------------------------------------------
\subsection{The US Senate}
%------------------------------------------------------
%
The techniques developed in this paper can be applied to political
voting.
Consider the 113$^{\text{th}}$ Congress, $1^{\text{st}}$ Session
(started January 2013) where there were 52 Democrats, 46 Republicans,
and 2 Independents in the Senate \cite{CongressByTheNumbers}.
To obtain cloture\footnote{``Cloture is a motion or process in
  parliamentary procedure aimed at bringing debate to a quick end.''
  \cite{ClotureWikipedia}} in the Senate 60 votes are sometimes
needed; this naturally leads to the $[60; 53, 45, 2]$ game.
We assume a voting structure in which the Democrats and Republican
teams have members each casting a single vote according to Equation
(\ref{eq:30}) and the Independents use ``random voting'' (are equally
likely to give 0 or 2 votes to any coalition).
That is:
\begin{equation}
\begin{aligned}
G_{\Dteam} 
&=   L_D \surround{\bigstrut (1-p_D)+ p_D    x }^{53}
+(1-L_D)\surround{\bigstrut    p_D + (1-p_D)x }^{53} \\
G_{\Rteam} 
&=   L_R \surround{\bigstrut (1-p_R)+ p_R    x }^{45}
+(1-L_R)\surround{\bigstrut    p_R + (1-p_R)x }^{45} \\
G_I 
&=\tfrac{1}{2}+\tfrac{1}{2}x^2
\end{aligned}
\label{eq:35}
\end{equation}
where $p_D$ (resp.~$p_R$) represents the probability that an
individual Democrat (resp.~Republican) votes the way their leader
desires as indicated by $L_D$ (resp.~$L_R$).

The Washington Post \cite{CongressVotesDatabase} lists the frequency
with which Democratic and Republican senators voted with their party
for the $112^{\text{th}}$ Congress.
For the Democrats the average value was 94\% while for the Republicans
it was 84\%; we refer to this as the \textit{cohesion} value.
For the $113^{\text{th}}$ Congress, we assume the values $p_R=0.94$
and $p_D=0.84$ for the Democratic and Republican cohesion.

When both the Democratic leader and the Republican leaders agree on an
issue then there is little contention.
Voting power becomes interesting when one team is in favor of an
action ($L=1$) and the other team is opposed ($L=0$).
Hence, consider two cases:
\begin{enumerate}

\item The Democratic leader wants to obtain cloture  ($L_D=1$) while the
  Republican leader is opposed to it ($L_R=0$).
  The Generalized Banzhaf power for the teams at the cohesion value
  are: Democrats 0.35, Republicans 0.35, Independents 0.30.

  It is somewhat surprising that the Democrats, Republicans, and
  Independents all have similar power, especially since the
  Independents have only two members!

\item The Republican leader wants to obtain cloture ($L_R=1$) while
  the Democratic leader is opposed to it ($L_D=0$).
  The Generalized Banzhaf power for the teams at the cohesion value
  are: Democrats 0.41, Republicans 0.31, Independents 0.28.

\end{enumerate}
The Generalized Banzhaf power for the three teams as $p_D$ and $p_R$
are varied, is shown in Figure~\ref{fig:5}.

Partial derivatives indicate how the Generalized Banzhaf values change
as the cohesion value changes.
At the cohesion point, $(p_R,p_D)=(0.94,0.84)$, we numerically
compute:
\begin{enumerate}

\item When ($L_D=1$) and  ($L_R=0$):
$\ds
\begin{aligned}[t]
\pdiff{ \beta'(\text{Dem}) }{p_D } =  0.04, \qquad
&\pdiff{ \beta'(\text{Dem}) }{p_R } = -0.36, 
\\
\pdiff{ \beta'(\text{Rep}) }{p_D } =  0.06,  \qquad
&\pdiff{ \beta'(\text{Rep}) }{p_R } = -0.37. 
\\
\end{aligned}$
\\
In this case, interestingly, both the Democrats and the Republicans
increase their power if either the Democratic cohesion increases or
the Republican cohesion decreases.

\item When  ($L_R=1$) and ($L_D=0$):
$\ds
\begin{aligned}[t]
\pdiff{ \beta'(\text{Dem}) }{p_D } =  -1.1, \qquad
&\pdiff{ \beta'(\text{Dem}) }{p_R } =  0.25, 
\\
\pdiff{ \beta'(\text{Rep}) }{p_D } =  0.44,  \qquad
&\pdiff{ \beta'(\text{Rep}) }{p_R } = -0.08. 
\\
\end{aligned}$
\\
In this case, the Democrats' power increases if either the Democratic
cohesion decreases or the Republican cohesion increases.
Just the opposite is true for the Republicans; their power increases
if either the Democratic cohesion increases or the Republican cohesion
decreases.

\end{enumerate}
In each of these cases the Republicans can adopt the same strategy to
increase their power: increase Democratic cohesion or decrease
Republican cohesion.

% 
% Ld = 1
% Lr = 0
% partial__power_D__pD =  0.04
% partial__power_D__pR = -0.36
% partial__power_R__pD =  0.06
% partial__power_R__pR = -0.37
%
% Ld=0
% Lr=1
% partial__power_D__pD = -1.1
% partial__power_D__pR =  0.25
% partial__power_R__pD =  0.44
% partial__power_R__pR = -0.08

\renewcommand\temp[2]{
  \parbox{\hsize}{                       \hfil
    \scalebox{0.4}{\includegraphics{#1}} \hfil
    \scalebox{0.4}{\includegraphics{#2}} \hfil
  }
}

\begin{figure}
\parbox{\hsize}{
  \hfil
  \hfil
  $(L_D=0, L_R=1)$
  \hfil
  \hfil
  \hfil
  $(L_R=1, L_D=0)$
  \hfil
}
\textbf{Dem}
\temp{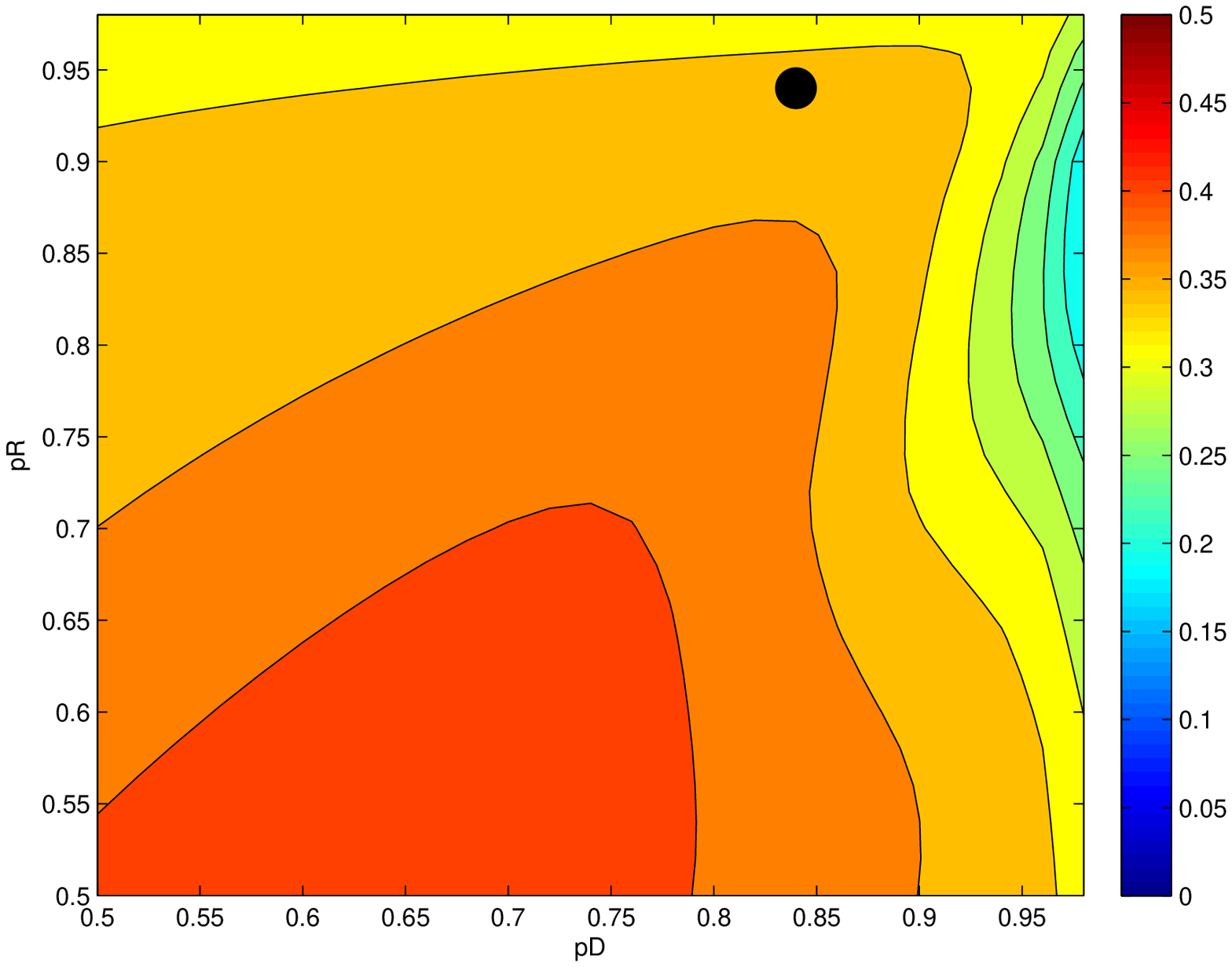}
     {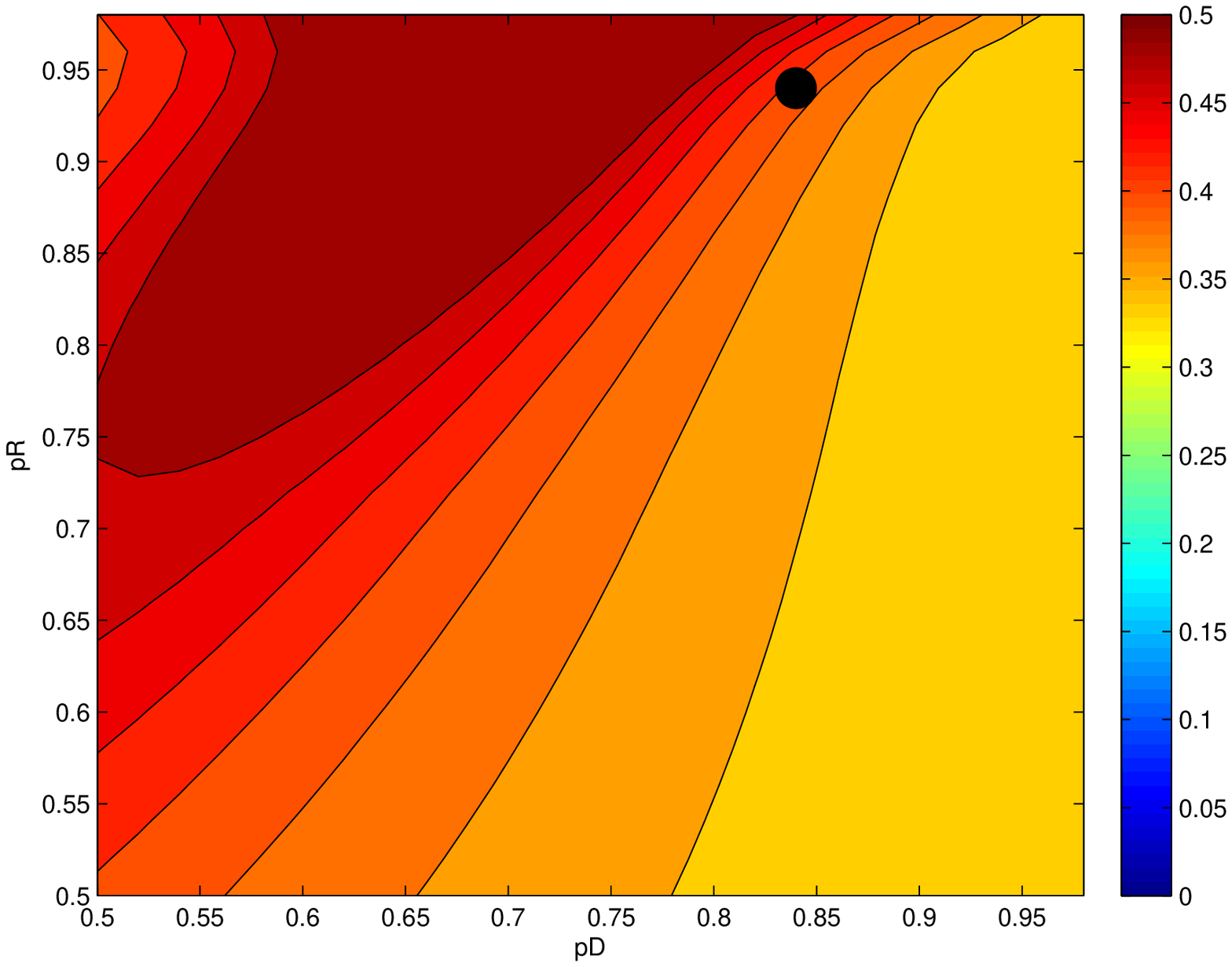}
\textbf{Rep}
\temp{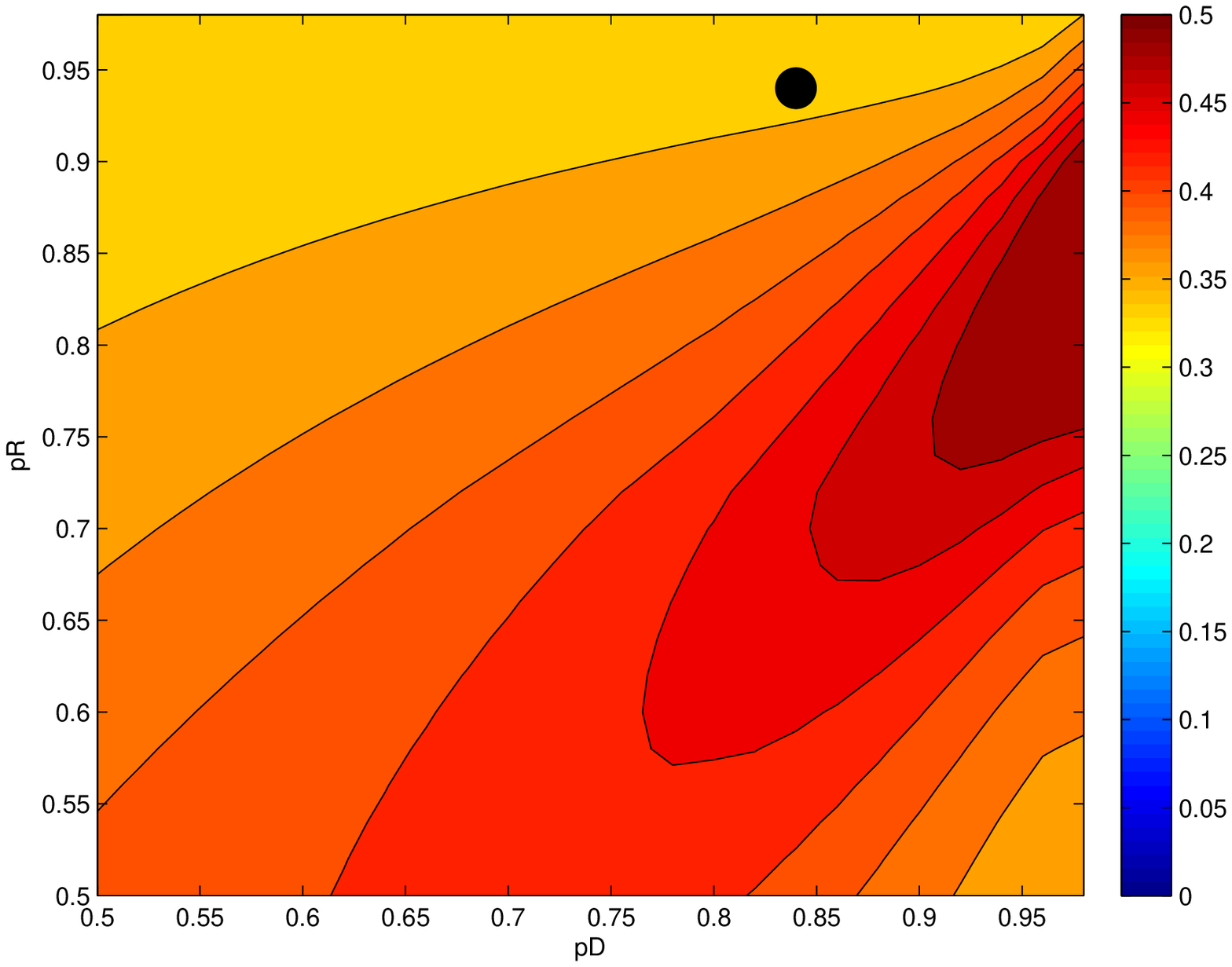}
     {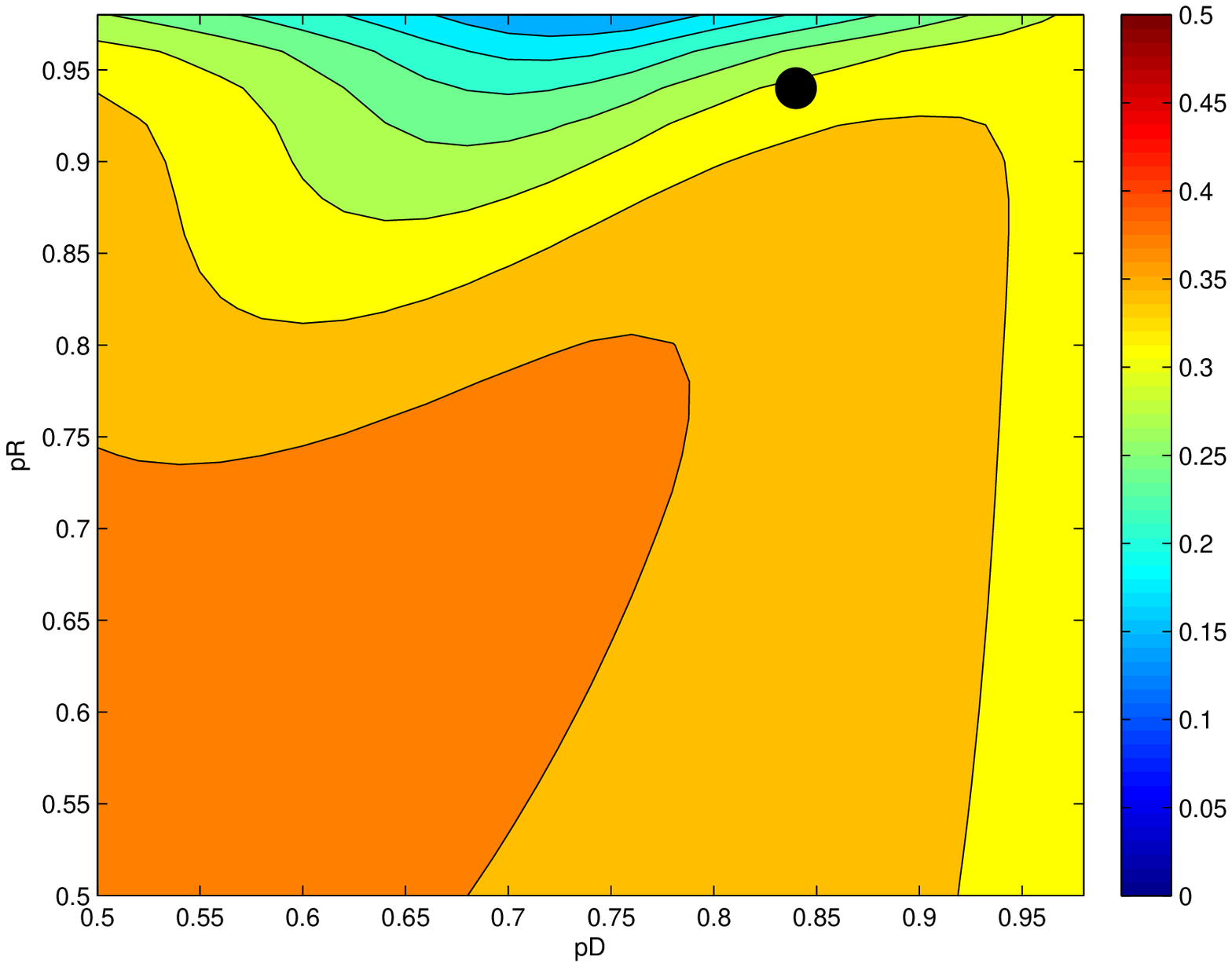}
\textbf{Ind}
\temp{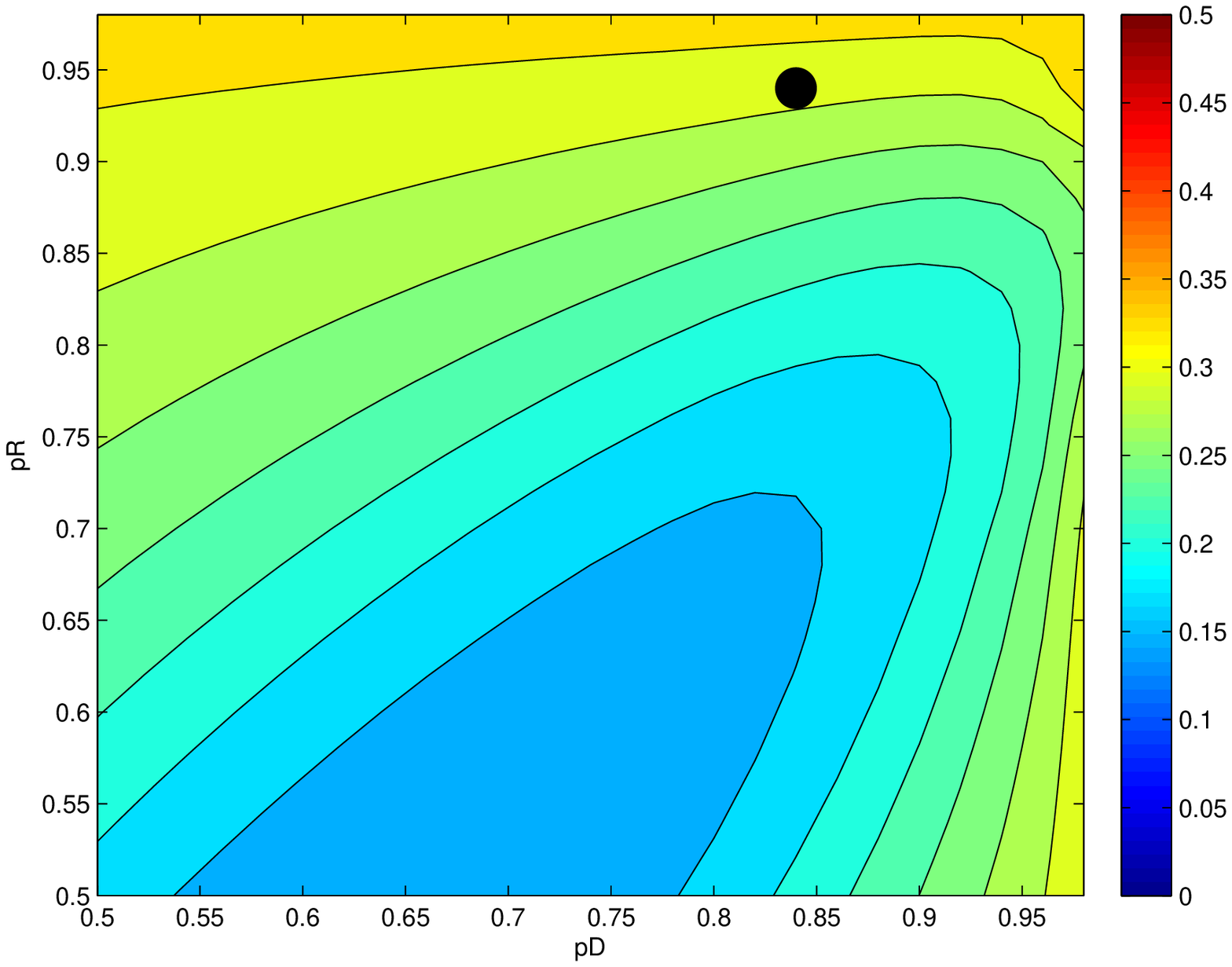}
     {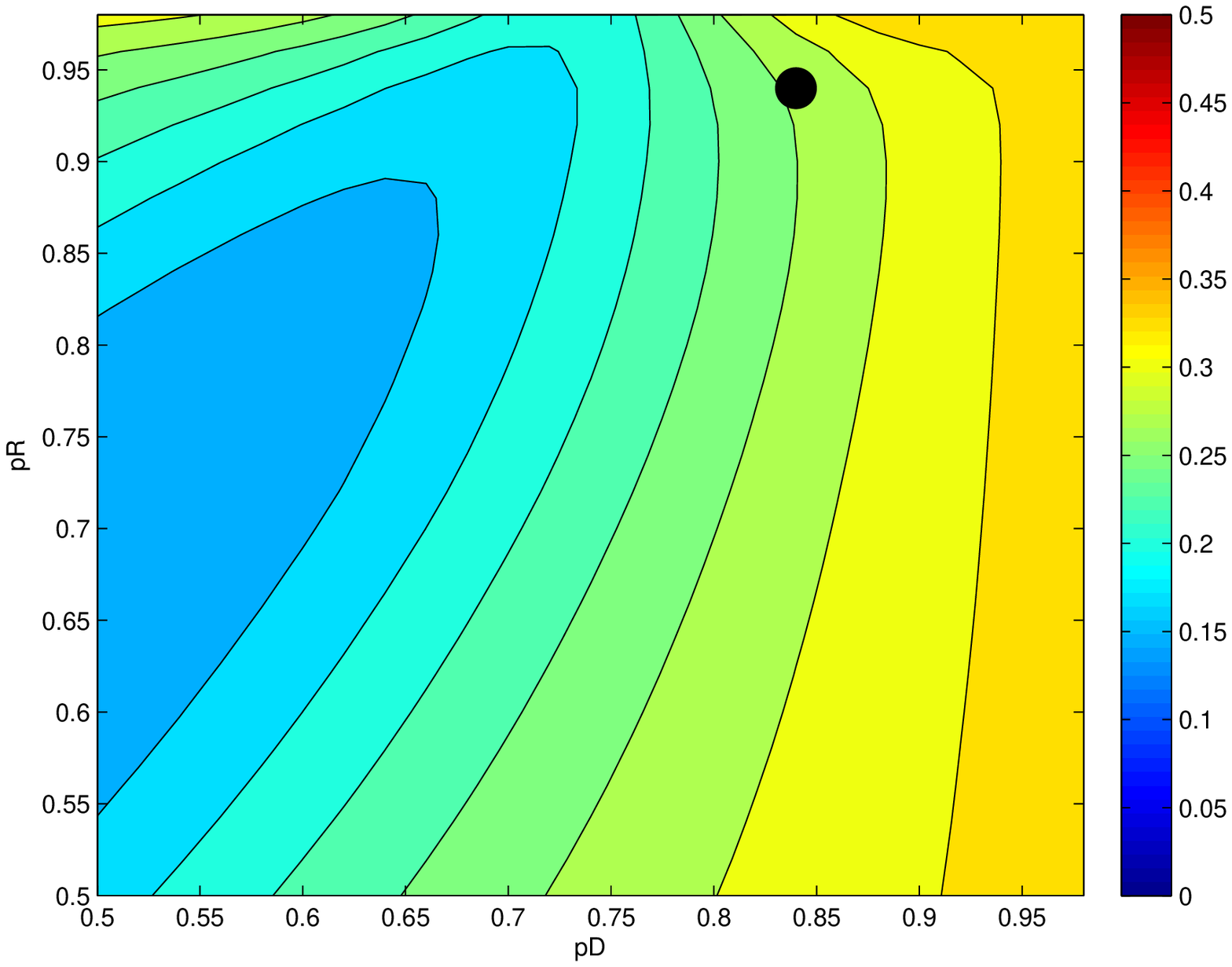}
\caption{The US Senate $[60,53,45,2]$ game. 
The left  column has $(L_D=0, L_R=1)$; 
the right column has $(L_R=1, L_D=0)$. 
The
top    row shows Democrats'    power;
middle row shows Republicans'  power;
bottom row shows Independents' power.
The dots show the cohesion point $(p_R,p_D)=(94\%, 84\%)$.
For each plot the horizontal axis is $p_D$ and the vertical axis is
$p_R$, both varying from $\frac12$ to 1.}
\label{fig:5}
\end{figure}

%------------------------------------------------------
\section{Summary}
%------------------------------------------------------
%
We have shown how to determine voting power when each player in a
weighted voting game has a ``voting structure'', a weighted generating
function representing probabilities of them contributing any number of
their votes to a coalition.
The resulting Generalized Banzhaf values can be computed with
polynomial arithmetic and reduce to the usual Banzhaf values when
random voting is used.

Voting structures can also be used to represent voter coalitions.
In this case each 
coalition who tries to influence the voting of each coalition member.
This model was applied to the US Senate to show who has (Democrats,
Republicans, or Independents) more power in attaining cloture.
When the Democrats are in favor of cloture and the Republicans are not
then, surprisingly, all three parties have similar power.

%------------------------------------------------------
% References
%------------------------------------------------------
\bibliographystyle{plain}
\bibliography{voting_power}

%------------------------------------------------------
\end{document}